\documentclass[aps,prr, twocolumn, grouppedaddress, nofootinbib]{revtex4-1}
\usepackage{amsmath,amssymb}
\usepackage{dsfont,yfonts}
\usepackage{tabularx}
\usepackage{multirow}
\usepackage{diagbox}
\newcolumntype{C}{>{\centering\arraybackslash}X}%
\usepackage{graphicx}
\usepackage{subfigure}
\usepackage[usenames,dvipsnames]{xcolor}
\usepackage{tikz}
\usepackage{pgffor}
\usepackage{verbatim}
\usepackage{bbm}
\usepackage{float}
\usepackage{mathrsfs}
\usepackage{xcolor,colortbl}
\usepackage[colorlinks, bookmarks=true,breaklinks=true,linkcolor=red, citecolor=blue, linktocpage=true, urlcolor=blue]{hyperref}
\newcommand{\Z}{\mathbb{Z}}

\newcommand{\Aut}{\rm Aut}

\newcommand{\modulo}[1]{\ ({\rm mod} \ #1)}
\makeatletter
\newcommand\xleftrightarrow[2][]{%
  \ext@arrow 9999{\longleftrightarrowfill@}{#1}{#2}}
\newcommand\longleftrightarrowfill@{%
  \arrowfill@\leftarrow\relbar\rightarrow}
\makeatother

\begin{document}

\title{Mirror anomaly in fermionic topological orders}

\author{Bin-Bin Mao}
\affiliation{Department of Physics and HKU-UCAS Joint Institute for Theoretical and Computational Physics, The University of Hong Kong, Pokfulam Road, Hong Kong, China}

\author{Chenjie Wang}
\email{cjwang@hku.hk}
\affiliation{Department of Physics and HKU-UCAS Joint Institute for Theoretical and Computational Physics, The University of Hong Kong, Pokfulam Road, Hong Kong, China}

\date{\today}

\begin{abstract}
We study general 2D fermionic topological orders enriched by the mirror symmetry with $\mathcal{M}^2=1$. It is known that certain mirror symmetry enriched fermionic topological orders (mirror SETs) are anomalous, in the sense that they cannot be realized in strict two dimensions but have to live on the surface of 3D topological crystalline superconductors. Mirror anomaly, or  equivalently 3D topological crystalline superconductor, has a $\Z_{16}$ classification. In this work, we derive an explicit expression, namely an \emph{anomaly indicator}, for the $\Z_{16}$ mirror anomaly for general fermionic mirror SETs. This derivation is based on the recently developed folding approach, originally proposed for bosonic topological orders. We generalize it to fermion systems. Through this approach, we establish a direct bulk-boundary correspondence between surface fermionic topological orders and 3D bulk topological crystalline superconductors. In addition, during the derivation, we obtain some general properties of fermionic topological orders as well as a few constraints on properties of fermionic mirror SETs.
\end{abstract}

\maketitle

%\tableofcontents

\section{Introduction}

The discovery of three-dimensional (3D) time-reversal symmetric topological insulators and topological superconductors has attracted tremendous attention in recent years, both experimentally and theoretically. \cite{hasan2010, Qi2011RMP} They are prominent examples of the so-called \emph{symmetry-protected topological} (SPT) phases, which are gapped short-range entangled states of matter that preserve certain global symmetries. \cite{chen2013,senthil2015} Different SPT phases can be obtained by varying symmetries and dimensions. 

In SPT phases, one of the key features is that the boundary cannot be trivially gapped. In one and two dimensions, it must be either gapless or breaking some symmetries\cite{CuPRB2006,WuPRL2006,ChenPRB2011a,LevinPRB2012a}. In three and higher dimensions, an additional possibility arises: the boundary can host a topological order that fully respects the symmetries\cite{vishwanath13}.  The relation between properties of the bulk and boundary is famously known as \emph{bulk-boundary correspondence} \cite{Witten1988,Wen1995}. This correspondence is many-to-one: different boundary states may correspond to the same SPT bulk.  For example, the surface of  a usual 3D topological insulator can host a single Dirac cone, as well as certain time-reversal symmetric non-Abelian topological orders\cite{metlitski15,bonderson13,wangc13b,chen14a,wangc14}. Symmetric surface topological orders have been studied for many other SPT phases recently\cite{vishwanath13,wangc13,burnell14,chen14,WangJPRX2018,fidkowski13,metlitski14,you14,WangPRX2016}.

A closely related concept is the so-called quantum anomaly, more precisely the \emph{'t Hooft anomaly}, in quantum field theories\cite{tHooftbook, KapustinPRL2014}. It is well known that some quantum field theory cannot be regularized (e.g., by a lattice realization) in a way that respects its global symmetries. Such symmetries are said to be anomalous. However, symmetric regularization is possible if the field theory is put on the boundary of a one dimension higher bulk, such that the bulk cancels the anomaly in the field theory, which is called anomaly inflow\cite{HarveyNPB1985,WittenRMP2016}. To have anomaly cancellation, it turns out that the bulk must be in certain SPT phase associated with the same symmetries. In fact, the 't Hooft anomalies in $d$ dimensional field theories with a symmetry group $G$ have a one-to-one correspondence to the $d+1$ dimensional SPT phases of the same group. This implies that anomaly is a topological property of field theories. If the field theory is anomaly-free, it means that the corresponding bulk is a trivial SPT. That being said, we see that establishing bulk-boundary correspondence for SPT phases is equivalent to identifying the 't Hooft anomaly in the quantum field theory on the boundary. In particular, when the boundary hosts a topological order, it is equivalent to identifying anomaly in a \emph{topological} quantum field theory. 

In this work, we perform a systematic study on quantum anomaly of the mirror symmetry $\mathcal{M}$ in 2D general fermionic topological orders. That is, we study the bulk-boundary correspondence for 3D fermionic SPT phases with mirror symmetry only, namely \emph{topological crystalline superconductors} (TCSCs), under the assumption that the surface is a mirror-symmetric topological order.  A few reasons that we focus on this case are as follows. First, quantum anomalies in bosonic topological orders in the presence of symmetries, i.e., symmetry-enriched topological (SET) phases, have been widely studied recently\cite{WenPRB2002,EssinPRB2013,MesarosPRB2013,TarantinoNJP2016,TeoAP2015,BarkeshliPRB2019,chen14, barkeshli2019relative}. However, the fermionic counterparts are much less studied. In fact, even in the absence of symmetries, fermionic topological orders are not completely understood.  Several recent developments on fermionic SETs can be found in Refs.~\onlinecite{FSET1, FSET2}.  Second,  there is a powerful method to study 't Hooft anomalies associated with internal unitary symmetries in SET phases: one couples the system to a background or dynamical gauge field and study topological properties of  symmetry defects in the gauged theory\cite{LevinPRB2012a,BarkeshliPRB2019,chen14,barkeshli2019relative}. If an SET is anomalous, topological properties of symmetry defects will display certain inconsistency. The anomaly can be read out from the inconsistency. However, for spatial symmetries (e.g., mirror symmetry) and anti-unitary symmetries (e.g., time-reversal symmetry),  there are no corresponding gauge fields. So, we need to look for other methods to analyze anomalies. Some recent works on anomalies of spatial and time-reversal symmetries can be found in Refs.~\onlinecite{QiPRL2015,HermelePRX2016,LakePRB2016, WangPRL2017,TachikawaPRL2017,Tachikawa2017b,ChengPRL2018,folding, Barkeshli2019b}. We will study mirror anomaly following the recently developed folding approach \cite{folding}, which will be discussed in detail in the main text. It was originally introduced for bosonic topological orders. In this work, we extend the folding approach to fermionic topological orders.

The mirror symmetry in those 3D TCSCs that we study satisfies $\mathcal{M}^2=1$. In the non-interacting limit, they have a $\Z$ classification, characterized by integer copies of Majorana cones on the surface. Strong interaction reduces the classification to $\Z_{16}$, which means that  mirror anomaly in 2D fermionic topological orders is classified by $\Z_{16}$ too\cite{wangc14, metlitski14, kitaev-z16,JWang2020}. We note that there are also 3D fermion systems with $\mathcal{M}^2=P_f$, where $P_f$ is the fermion parity  --- a symmetry that must be preserved in all fermion systems. However, there are no non-trivial 3D TCSCs in this case, i.e., all 2D fermionic topological orders with $\mathcal{M}^2=P_f$ are free from 't Hooft anomaly of the mirror symmetry. Therefore, we only study the case that $\mathcal{M}^2=1$.

We show that mirror anomaly in any 2D fermionic topological orders can be computed through the following explicit formula
\begin{equation}
\eta_{\mathcal{M}} = \frac{1}{\sqrt{2}D_\mathcal{C}} \sum_{a\in \mathcal{C}} d_a \theta_a \mu_a,
\label{eq:etam}
\end{equation}
where $\eta_{\mathcal{M}} = 1, e^{i \pi/8},\dots, e^{i 15\pi/8}$, with these sixteen values indicating the $\Z_{16}$ classification of mirror anomaly. We call the quantity $\eta_\mathcal{M}$ an \emph{anomaly indicator}.  A mirror symmetry enriched fermionic topological order is anomaly-free if and only if $\eta_{\mathcal{M}}=1$. The summation in \eqref{eq:etam} is over all anyons $a$ in a fermionic topological order denoted by $\mathcal{C}$. The quantity $d_a$ is the quantum dimension of anyon $a$, $D_\mathcal{C} = \sqrt{\sum_a d_a^2}$ is the total quantum dimension, $\theta_a$ is the topological spin of $a$, and $\mu_a =0, \pm 1$ is a quantity that characterizes mirror symmetry fractionalization (see definitions in Sec.~\ref{sec:mset}). 

The main results of this work are the explicit derivation of the anomaly indicator $\eta_{\mathcal{M}}$ in \eqref{eq:etam}  and generalization of the folding approach~\cite{folding} to fermionic topological orders. During the derivation of $\eta_{\mathcal{M}}$, we also obtain some general properties of fermionic topological orders. We will see that our derivation is a direct establishment of bulk-boundary correspondence. In fact, $\eta_\mathcal{M}$ can be interpreted as a bulk quantity describing the bulk of 3D TCSCs, while all quantities on the right-hand side of \eqref{eq:etam} describe properties of the surface topological order. Hence, Eq.~\eqref{eq:etam} not only provides a simple way to compute mirror anomaly, but also is a \emph{quantitative} bulk-boundary correspondence.

We point out that the expression \eqref{eq:etam} is totally expected.  Reference \onlinecite{WangPRL2017} conjectured an anomaly indicator $\eta_\mathcal{T}$ for  time-reversal symmetric fermionic topological orders, with $\mathcal{T}^2=P_f$. Its explicit expression is given by  
\begin{equation}
\eta_{\mathcal{T}} = \frac{1}{\sqrt{2}D_\mathcal{C}} \sum_{a\in\mathcal{C}} d_a \theta_a \tilde{\mathcal{T}}^2_a,
\label{eq:etaT}
\end{equation}
where $\tilde{\mathcal{T}}^2_a = 0,\pm 1$ describes time-reversal properties of anyon $a$ (e.g., whether it carries a Kramers degeneracy, etc; see Ref.~\onlinecite{WangPRL2017} for details). This conjecture was proved in Ref.~\onlinecite{TachikawaPRL2017} through an analysis based on continuum topological quantum field theory. It is obvious that \eqref{eq:etam} and \eqref{eq:etaT} are very similar. This similarity is expected from the topological crystalline equivalence principle\cite{TCEP}, which states that classifications of SPT/SET phases are equivalent for internal (such as time-reversal) and crystalline (such as mirror) symmetries. It is also expected from the similarity between time-reversal and mirror anomalies in the bosonic systems\cite{WangPRL2017, folding}.  We believe the proof of Ref.~\onlinecite{TachikawaPRL2017} can be easily adjusted to mirror-symmetric topological orders, as in the continuum time-reversal and mirror reflection are related by Lorentz transformations. Compared to the proof of Ref.~\onlinecite{TachikawaPRL2017}, the novelty of our derivation of \eqref{eq:etam} includes two aspects: (i) it establishes a direct bulk-boundary correspondence and (ii) it uses algebraic properties of fermionic SETs,  rather than a field theoretical analysis.

\begin{figure*}[t]
\centering
\includegraphics[scale=1]{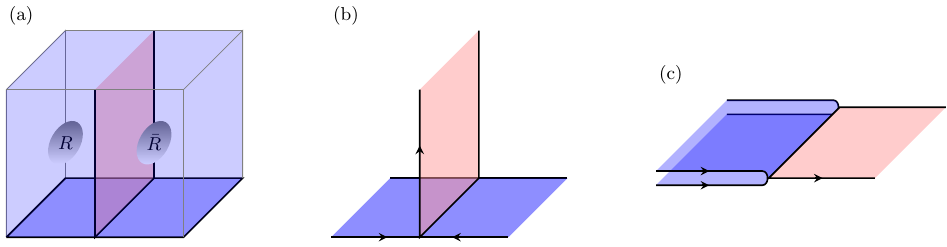}
\caption{Dimensional reduction and folding. (a) Short-range entanglement (light blue) in the disjoint union $R\bigcup \bar{R}$ can be removed by applying a mirror-symmetric local unitary transformation $\mathcal{U}\cdot \mathcal{M}^{-1}\mathcal{U}\mathcal{M}$. (b) By enlarging $R$ and $\bar{R}$, only entanglement on and near the mirror plane (light red) and the surface  (dark blue) remains, giving rise to an inverted-T-like junction decoupled from other degrees of freedom. The two wings of the inverted-T-like junction are mirror images of each other, hosting topological orders with opposite chiralities. Arrows represent chiralities. (c) We fold the two wings and obtain a double-layer system. The domain wall between the double-layer system and mirror plane is gapped.  }
\label{fig:folding}
\end{figure*}

The rest of the paper is organized as follows. In Sec.~\ref{sec:dimreda}, we review the dimensional reduction description of 3D TCSCs, introduce the folding approach, and set up the main system in Fig.~\ref{fig:folding}(c) for the derivation of Eq.~\eqref{eq:etam}. In Sec.~\ref{sec:eta}, we give a bulk interpretation of $\eta_\mathcal{M}$. In Sec.~\ref{sec:double}, we describe properties of mirror symmetric fermionic SETs as well as several topological orders that will occur in the folding approach. Next, we analyze properties of the gapped domain wall in Fig.~\ref{fig:folding}(c) through anyon condensation theory in Sec.~\ref{sec:anyoncon}. In Sec.~\ref{sec:indicator}, we derive the expression \eqref{eq:etam} of the anomaly indicator. We describe several explicit examples in Sec.~\ref{sec:examples}. We conclude in Sec.~\ref{sec:conclusion}. In Appendix \ref{sec:app_B_prop}, we discuss some general properties of fermionic topological orders after the fermion parity is gauged.

\section{General picture}
\label{sec:dimred}

In this section, we discuss the general physical picture behind the folding approach\cite{folding} in the context of 3D TCSCs. In particular, we give a bulk interpretation of the anomaly indicator $\eta_{\mathcal{M}}$.

\subsection{Dimensional reduction and folding}
\label{sec:dimreda}

3D TCSCs are topologically non-trivial because they are short-range entangled and the entanglement cannot be removed by finite-depth local unitary transformations \cite{ChenPRB2010} (LUTs) in a fully symmetric way without closing the energy gap.  Nevertheless, if we ignore the mirror symmetry, entanglement can be fully removed by finite-depth LUTs and the ground state can be smoothly deformed into a trivial state. Below we review the physics of 3D TCSCs, following the dimensional reduction approach of Ref.~\onlinecite{SongPRX2017}. 

Imagine a 3D TCSC with the mirror plane in the middle [Fig.~\ref{fig:folding}(a)]. Let $R$ be a region inside the bulk on the left-hand side of the mirror plane, and $\bar{R}$ be its mirror image. The two regions  do not overlap. Since the bulk is short-range entangled, we can apply a LUT $\mathcal{U}$ on region $R$ and remove all the entanglement inside. To make it symmetric, we also apply $\mathcal{M}^{-1}\mathcal{U}\mathcal{M}$ on region $\bar{R}$, where $\mathcal{M}$ denotes the unitary operator of mirror symmetry. Indeed, the product $\mathcal{U}\cdot \mathcal{M}^{-1}\mathcal{U}\mathcal{M}$ respects the mirror symmetry and removes the short-range entanglement in the regions $R$ and $\bar{R}$. One can continue to remove entanglement by enlarging the regions $R$ and $\bar{R}$. Finally, all entanglement in the bulk is removed, only except near the mirror plane [Fig.~\ref{fig:folding}(b)]. On and near the mirror plane, $R$ and $\bar{R}$ overlap, so the above LUT does not work. This leads to a dimensional reduction: the bulk physics is captured by the effective 2D system on (and near) the mirror plane. Moreover, the mirror symmetry is effectively an \emph{internal} $\Z_2$ symmetry on the mirror plane, making it easier to be analyzed. Accordingly, the effective system is a 2D invertible topological order (iTO) with an internal $\Z_2$ symmetry. We will discuss properties of the effective system below in Sec.~\ref{sec:eta}.

Next we consider a TCSC in the presence of a surface. We assume that the surface is mirror-symmetric and topologically ordered throughout the paper. In this case, the surface is long-range entangled, i.e.,  entanglement cannot be removed even in the absence of any symmetry. Nevertheless, we can still apply mirror-symmetric LUTs as above, both in the bulk and on the surface. While not being able to turn the surface into the trivial state, we can deform distinct mirror SETs into \emph{almost} the same state, for a fixed surface topological order. The only exception is near the intersection line of the mirror plane and the surface, where distinct mirror SETs can not be deformed into the same. Then, we obtain an inverted-T-like junction [Fig.~\ref{fig:folding}(b)], which contains all the entanglement and decouples from other degrees of freedom. We see that all information of mirror SETs is transformed onto the intersection line between the surface and the mirror plane. Accordingly, the intersection line is the key to mirror SETs. It is worth mentioning that the left and right wings of the inverted-T-like junction have opposite chiralities, as they are mirror partners of each other. 

Therefore, it is enough to consider the inverted-T-like junction only. Ref.~\onlinecite{folding} proposes to fold the two wings of the junction and turn it into Fig.~\ref{fig:folding}(c). Then, the left-hand side is a double-layer topological order, and the right-hand side is the original mirror plane hosting a $\mathbb{Z}_2$ symmetric invertible topological order. The whole system in Fig.~\ref{fig:folding}(c), including the domain wall, is energetically gapped. Upon folding, the mirror symmetry, which originally exchanges the two surface wings,  now becomes a layer-exchange symmetry on the double-layer system.  Accordingly, the whole system in Fig.~\ref{fig:folding}(c) also has an internal $\Z_2$ symmetry. From now on, we will denote this internal symmetry  as $\Z_2^{\rm ex}$ for clarity (for the iTO, we still refer it as $\Z_2$ sometimes when it does not cause any confusion). The advantage of turning the mirror symmetry into an internal symmetry is that we can now gauge it and study mirror-SETs as if they are internal SETs.

To sum up, through dimensional reduction and folding, we obtain the main setup, shown in Fig.~\ref{fig:folding}(c). It contains three parts: (i) the double-layer topological order on the left-hand side, (ii) the invertible topological order on the right-hand side, and (iii) a gapped domain between them. All parts are symmetric under $\Z_2^{\rm ex}$. The double-layer topological order is \emph{universal}, in the sense that it is the same for all mirror SETs, given a fixed intrinsic topological order\cite{folding}. Properties of the double-layer system will be discussed in detail in Sec.~\ref{sec:double}. As discussed above, all information about mirror SETs are  encoded at the domain wall. In Sec.~\ref{sec:anyoncon}, we will use anyon condensation theory to study the domain wall. Properties of the domain wall are the keys to understand mirror symmetry fractionalization. Finally, the invertible topological order corresponds to the bulk TCSC and the anomaly of the surface SETs. We discuss the invertible topological order(iTO) below in Sec.~\ref{sec:eta}.

Two comments are in order. First, for purely 2D mirror SETs, we can think of them as living on the surface of a trivial mirror-symmetric 3D system. Then, the above analysis still holds. The effective 2D system on the mirror plane must be in a trivial phase. Second, in this work, we have focused the mirror symmetry with $\mathcal{M}^2=1$. When $\mathcal{M}^2= P_f$, it is known that there is no non-trivial 3D TCSCs. This can also be seen from dimensional reduction. The bulk will be reduced to a 2D system with $\Z_4^f$ internal symmetry, which does not support any nontrivial iTO after taking care of adjoining operations.\cite{SongPRX2017,WangPRB2016}

\begin{figure}[b]
\centering
\includegraphics[scale=1]{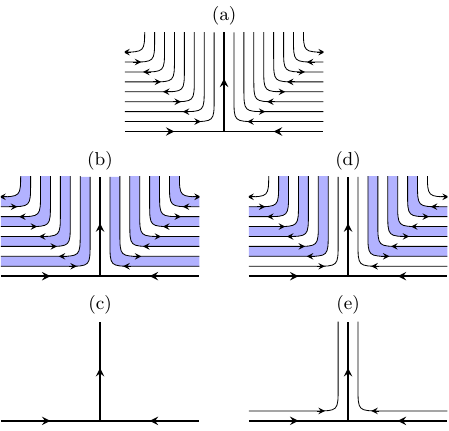}
\caption{Adjoining operation by placing $p_x\pm ip_y$ superconductors on the two sides of the mirror plane to the inverted-T-like junction in Fig.~\ref{fig:folding}(b). For simplicity, only front views of the 3D system are shown. Arrows correspond to chirality. Two lines connected by a blue stripe correspond to a pair of $p_x+ip_y$ and $p_x-ip_y$ superconductors that can be removed by finite depth LUTs. Existence of the two paths, (a)-(b)-(c) and (a)-(d)-(e), shows that the states in (c) and (e) are topologically equivalent. We see that (e) has two extra $p_x + ip_y$ superconductors on the two sides of the mirror plane, compared to (c).}
\label{fig:adjoining}
\end{figure}

\subsection{Bulk interpretation of $\eta_{\mathcal{M}}$}
\label{sec:eta}
As discussed in the introduction, $\eta_\mathcal{M}$ in Eq.~\eqref{eq:etam} is a bulk quantity that characterizes 3D TCSCs. Ref.~\onlinecite{TachikawaPRL2017} gives an interpretation  of $\eta_\mathcal{M}$ through topological quantum field theories that are defined on (3+1)D un-oriented space-time manifolds.  Here, we give an alternative interpretation through the invertible topological order in Fig.~\ref{fig:folding}(c). More explicitly, we give the quantitative definition in Eq.~\eqref{eq:etam-def}.

Before giving our interpretation, we discuss a subtlety on the equivalence between $\Z_2$-symmetric iTOs and 3D TCSCs.  According to Ref.~\onlinecite{GuLevin2014}, 2D iTOs with an internal $\Z_2$ symmetry (an overall $\Z_2^f\times \Z_2$ symmetry group) are classified by $\Z\times \Z_8$.  On the other hand, 3D TCSCs have a $\Z_{16}$ classification. To resolve the mismatch, one needs to consider the so-called \emph{adjoining operation} \cite{SongPRX2017}, shown in Fig.~\ref{fig:adjoining}. It shows that through proper LUTs,  placing two $p_x+ip_y$ superconductors (or $p_x-ip_y$ superconductors) symmetrically on the two sides of the mirror plane is topologically equivalent to doing nothing. Accordingly, under the adjoining operation, certain non-trivial phases when viewed as authentic 2D iTOs become trivial when viewed as effective 2D phases that are dimensionally reduced from 3D TCSCs. That is, the classification is reduced.

To see why it is reduced to $\Z_{16}$, we take a close look at the $\Z\times \Z_8$ classification of the $\Z_2$-symmetric iTOs. The root state for the $\Z$ classification is the $p_x+i p_y$ superconductor, on which the $\Z_2$ symmetry acts trivially. The root state for the $\Z_8$ classification is a non-chiral SPT state, which can be obtained by simply stacking a $p_x+ip_y$  superconductor with a $p_x-ip_y$ superconductor. The edge of the second root state can be described by a pair of counter-propagating chiral Majorana fermions
\begin{equation}
H = i\psi_L \partial_x \psi_L  - i\psi_R \partial_x \psi_R,
\end{equation}
where $\psi_\sigma^\dag = \psi_\sigma$, with $\sigma = L,R$. Here, $L$ and $R$ stand for left- and right-moving modes. The $\Z_2$ symmetry, which we denote as $\mathbf{x}$, acts on the fermions as follows
\begin{equation}
\mathbf{x}: \ \psi_L \rightarrow \psi_L, \quad \psi_R \rightarrow -\psi_R.
\end{equation}
That is, $\psi_L$ is neutral while $\psi_R$ is charged. A general phase in the $\Z\times \Z_8$ classification can then be labeled by an integer pair $(\mu_1, \mu_2)$, where $\mu_2 $ is defined only modulo 8. The two integers count the numbers of root states in a general phase. Since $p_x + ip_y$ has a chiral central charge $c =1/2$, a general phase indexed by $(\mu_1, \mu_2)$ shall have a central charge $c=\mu_1/2$.

Now we analyze how precisely the adjoining operation modifies the classification of 3D TCSCs. To do that, we imagine the mirror plane has an edge. Under the adjoining operation, the edge acquires two additional right-moving chiral Majorana fermions, $\psi_{R1}$ and $\psi_{R2}$. Under the mirror symmetry action, we have
\begin{equation}
\mathcal{M}:\  \psi_{R1} \leftrightarrow \psi_{R2}.
\end{equation}
Let us perform a unitary transformation by defining
\begin{align}
\tilde{\psi}_{R_1} = \frac{1}{\sqrt{2}}(\psi_{R1} + \psi_{R2}),\nonumber\\
\tilde{\psi}_{R_2} = \frac{1}{\sqrt{2}}(\psi_{R1} - \psi_{R2}).
\end{align}
The two new fermions transform as $\tilde{\psi}_{R1} \rightarrow \tilde{\psi}_{R1}$, and $\tilde{\psi}_{R1} \rightarrow - \tilde{\psi}_{R1}$. Then, it is not hard to see that this edge corresponds to $(\mu_1, \mu_2) = (2, 1)$. Accordingly, $(\mu_1, \mu_2) = (2, 1)$ corresponds to a trivial state in the classification of 3D TCSCs. Let us define the index
\begin{equation}
\nu = \mu_1 - 2\mu_2.
\label{eq:nu}
\end{equation}
It is invariant under adjoining operations. One can see that $\nu$ is distinct up to modulo 16, giving rise to a $\Z_{16}$ classification of 3D TCSCs. 

With the above understanding, we now define $\eta_{\mathcal{M}}$ through properties of 2D $\Z_2$-symmetric iTOs. According to Ref.~\onlinecite{levin2012}, SPTs and iTOs with internal symmetries can be studied by gauging the symmetries and analyzing the braiding statistics between the symmetry fluxes. In our case, the total symmetry is $\Z_2^f\times \Z_2$. The states in the $\Z$ classification are characterized by the central charge $c=\mu_1/2$ of its edge theory.  The topological spin of any fermion parity flux $w$ is 
\begin{equation}
\theta_{w} = e^{i2\pi\mu_1/16}.
\label{eq:pfmu1}
\end{equation}
Note that there may be many fermion parity fluxes, but all have the same topological spin. The states in the $\Z_8$ classification are characterized by 
\begin{equation}
\theta_{x}^2 = e^{i2\pi \mu_2/8},
\label{eq:gmu2}
\end{equation}
where $x$ denotes a $\Z_2$ flux. The topological spin $\theta_x$ itself is not a topological invariant because different $\Z_2$ fluxes have different topological spins, but they only differ by a minus sign.\cite{GuLevin2014,WangPRB2017} Accordingly, the squared quantity $(\theta_x)^2$ is insensitive to the choice of flux. It is an topological invariant. More detailed discussion on gauged iTOs can be found in Sec.~\ref{sec:z2z2ito}.

Finally we consider the index $\nu$ that labels the $\Z_{16}$ classification of 3D TCSCs. We define $\eta_{\mathcal{M}} = e^{i2\pi \nu/16}$. Then, with the equations \eqref{eq:nu}, \eqref{eq:pfmu1} and \eqref{eq:gmu2},  we obtain the following relation,
\begin{equation}
\eta_{\mathcal{M}} = \theta_{w} (\theta_x^*)^2.
\label{eq:etam-def}
\end{equation}
That is, the quantity $\eta_{\mathcal{M}}$ can be interpreted as a special combination of the topological spins of a fermion parity flux and a $\Z_2$ flux. Equation \eqref{eq:etam-def} serves as the definition of $\eta_\mathcal{M}$. It is the starting point to derive the formula \eqref{eq:etam}.

\section{Double-layer topological order}
\label{sec:double}
In this section, we study properties of the double-layer topological order in Fig.~\ref{fig:folding}(c). The main strategy is to gauge the $\Z_2^f\times \Z_2^{\rm ex}$ symmetry and study properties of the symmetry fluxes.

\subsection{Mirror SETs}
\label{sec:mset}

To start, we describe several quantities that characterize the mirror-SETs before folding. Quantities for both topological and symmetry properties will be discussed.

\subsubsection{Topological properties}

General algebraic theory of anyons can be found in Ref.~\onlinecite{kitaev2006}. Here, we mention a few basic properties. Let us denote the surface topological order by $\mathcal{C}$, and let $|\mathcal{C}|$ be the number of anyons it contains.  We will use $\mathbbm{1},a, b, \dots$ to denote the anyons, where $\mathbbm{1}$ is the vacuum anyon representing local bosonic excitations. Anyons satisfy the fusion rules $a\times b = \sum_c N_{ab}^c c$, where $N_{ab}^c$ is called the fusion coefficient. In particular, for each anyon $a$, there exists a unique anti-particle $\bar{a}$, such that $N_{a\bar{a}}^{\mathbbm{1}} =1$. From the fusion and braiding properties of anyons, one can also define other topological quantities, including quantum dimension $d_a$, topological spin $\theta_a$, the matrices $T$ and $S$, etc. When $d_a=1$, anyon $a$ is said to be Abelian; otherwise, $a$ is non-Abelian. The quantum dimension satisfies $d_ad_b = \sum_c N_{ab}^c d_c$. The quantity $\mathcal{D}_\mathcal{C} = \sqrt{\sum_a d_a^2} $ is called the total quantum dimension of $\mathcal{C}$. In the case of Abelian anyons,  topological spin $\theta_a$ is equivalent to the phase obtained by exchanging two identical $a$ anyons; for non-Abelian anyons, $\theta_a$ has only an algebraic definition (see Ref.~\onlinecite{kitaev2006}). We have $\theta_{\bar a} =\theta_a$. The matrices $T$ and $S$ are of size $|\mathcal{C}|\times |\mathcal{C}|$. They are defined by
\begin{align}
T_{a,b} & = \theta_a \delta_{a,b} ,\nonumber\\
S_{a,b} & = \frac{1}{D_\mathcal{C}} \sum_c N_{a\bar b}^c\frac{\theta_c}{\theta_a\theta_b} d_c .
\end{align}
For topological orders whose constituent particles are bosons, an important property is that $S$ is a unitary matrix, i.e., $S^\dag S =1$. This property is called the \emph{modularity} of bosonic topological orders. Modularity guarantees that $\mathbbm{1}$ is the only particle that has trivial mutual braiding statistics, or is ``transparent'', with respect to all anyons in the system. Mathematically speaking, a bosonic topological order is equivalent to a unitary modular tensor category (UMTC). 

In our case, the constituent particles are fermions. The significant difference in fermion systems is that in addition to $\mathbbm{1}$, there always exists another local excitation, the fermion $f$. Like $\mathbbm{1}$, the fermion $f$ is also transparent to all anyons. That is,
\begin{equation}
M_{f, a} = 1,
\label{eq:monodromy1}
\end{equation} 
for all $a\in \mathcal{C}$, where $M_{f,a}$ denotes the mutual statistics between $f$ and $a$. In general, mutual statistics depends on the fusion channel, but not if one of the two anyons is Abelian. The existence of $f$ makes the $S$ matrix in fermionic topological orders not unitary and thereby fermionic topological orders are not modular. Mathematically speaking, fermionic topological orders are described by unitary pre-modular tensor categories \cite{Drinfeld2010, Bruillard2017,LanPRB2016}.  Another feature of fermionic topological orders is that all anyons come in pairs, $a$ and $af$, where $a\times f = af$. That is, 
\begin{equation}
\mathcal{C} = \{\underbrace{\mathbbm{1}, f}_{[\mathbbm{1}]}, \underbrace{a, af}_{[a]},\underbrace{b, bf}_{[b]}, \dots\},
\end{equation}
where we use $[a]$ to denote the pair $\{a, af\}$ for later convenience. Note that $[a]=[af]$. The topological spins satisfy $\theta_{af} = -\theta_a$.  In some special cases, we can separate a fermionic topological order $\mathcal{C}$ into a ``direct product'' of $\{1, f\}$ and bosonic topological order $\mathcal{C}_{bto}$, in the sense that all fusion and braiding properties can be separated. We will denote it as $\mathcal{C} = \{1,f\}\boxtimes \mathcal{C}_{bto}$.  One may physically think of the symbol ``$\boxtimes$'' as stacking.  Accordingly, $\mathcal{C}$ is a simple stack of the trivial fermionic topological order and a bosonic topological order $\mathcal{C}_{bto}$.  However, generally speaking, such separation may not be possible. We will refer those topological orders that cannot be separated as \emph{intrinsically fermionic} topological orders.

%For example, the $SO(3)_3$ Chern-Simons theory supports four anyons $\{1, f, s, s_f$. One of the fusion rules is that $s\times s = 1 + s + s_f$, which is enough to show that $SO(3)_3$ is intrinsic. 

\subsubsection{Symmetry properties}
\label{sec:symmetry}

The surface topological order of 3D TCSCs preserves the mirror symmetry $\mathcal{M}$, which satisfies $\mathcal{M}^2 =1$. In general, the interplay between symmetry and topological order in SETs includes three layers of data\cite{BarkeshliPRB2019}: (i) anyon permutation by symmetries, (ii) symmetry fractionalization on anyons, and (iii) stacking of 2D SPT phases. Below we describe them one by one for mirror SETs.  

First, anyon permutation by the mirror symmetry is described by a group homomorphism $\rho: \Z_2^\mathcal{M} \rightarrow \Aut^*(\mathcal{C})$, where $\Z_2^\mathcal{M} = \{1, \mathcal{M}\}$ is the mirror symmetry group and $\Aut^*(\mathcal{C})$ is the group of autoequivalences and anti-autoequivalences of the topological order $\mathcal{C}$. An autoequivalence is a one-to-one map from $\mathcal{C}$ to itself such that all the fusion and braiding quantities are preserved. An anti-autoequivalence is a one-to-one map from $\mathcal{C}$ to itself such that all the fusion and braiding quantities are preserved with complex conjugation. The mirror symmetry $\mathcal{M}$ reserves the spatial orientation, so $\rho(\mathcal{M})$ must be an anti-autoequivalence. At the same time, $\rho(1)$ must be the trivial autoequivalence, i.e., $\rho(1) = \mathbbm{1}$. For simplicity, we will use the short-hand notation $\rho(\mathcal{M}) = \rho_m$. Here, we list a few useful properties that $\rho_m$ should satisfy. To respect $\mathcal{M}^2 =1$, we have $\rho_m^2 = \mathbbm{1}$. That is, permuting any anyon $a$ twice is trivial, $\rho_m(\rho_m(a))=a$. As an anti-autoequivalence, we have $N_{\rho_m(a)\rho_m(b)}^{\rho_m(c)} = N_{ab}^c$ and $\theta_{\rho_m(a)} = \theta_a^*$.  In addition, $\rho_m(\bar a) = \overline{\rho_m(a)}$. We will see later that it is convenient to define an anti-autoequivalence $\bar\rho_m$ by $\bar\rho_m(a) = \overline{\rho_m(a)}$. In fermionic topological orders, the two local excitations cannot be permuted by mirror symmetry, so we must have $\rho_m(\mathbbm{1}) = \mathbbm{1}$ and $\rho_m(f) = f$. 

\begin{figure}
\centering
\includegraphics[scale=1]{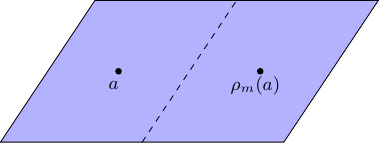}
\caption{Defining mirror symmetry fractionalization through the mirror eigenvalue $\mu_a$ of a mirror-symmetric two-anyon state $|a,\rho_m(a)\rangle$. Such two-anyon state is physically possible only if $a$ and $\rho_m(a)$ fuse to local excitations, i.e., $\mathbbm{1}$ and $f$.}
\label{fig:mu}
\end{figure}

Second, to characterize fractionalization of the mirror symmetry, one can consider a two-anyon state $|a,\rho_m(a)\rangle$. The two anyons $a$ and $\rho_m(a)$ are symmetrically located on the two sides of the mirror axis (Fig.~\ref{fig:mu}).  This state can be made mirror symmetric by locally adjusting the wave function near $a$ and $\rho_m(a)$. So, it has a well-defined mirror eigenvalue $\mu_a = \pm 1$:
\begin{equation}
\mathcal{M} |a,\rho_m(a)\rangle = \mu_a |a, \rho_m(a)\rangle.
\label{eq:mirroreig}
\end{equation}
It was shown in Ref.~\onlinecite{ZaletelPRB2017} that $\mu_a$ cannot be changed in a given SET and so it is topologically robust. Note that the state $|a, \rho_m(a)\rangle$ is physically possible only if the  two anyons fuse to local excitations, i.e., $\mathbbm{1}$ and $f$. That is, we have either $\rho_m(a) =\bar a$ or $\rho_m(a) =\bar a f$. For later convenience, we define
\begin{equation}
\xi_a =\left\{
\begin{array}{ll}
1, &\quad \text{if } \rho_m(a)=\bar a,\\
-1, & \quad \text{if }  \rho_m(a) = \bar a f,\\
0, &\quad \text{otherwise},
\end{array}
\right. 
\label{eq:xi}
\end{equation}
and
\begin{equation}
\mu_a =0, \quad \text{if $\xi_a=0$}.
\label{eq:mirroreig2}
\end{equation}
Since $\theta_{\rho_m(a)} = \theta_a^*$, we must have $\theta_a = \pm 1$ if  $\rho_m(a) =\bar a$, and  $\theta_a = \pm i$ if  $\rho_m(a) =\bar a f$. The quantity $\mu_a=\pm 1$ for $\rho_m(a)= \bar{a}f$ is analogous to the ``$\mathcal{T}^2_a= \pm i$'' fermionic Kramers' degeneracy, first discussed in Refs.~\onlinecite{fidkowski13,metlitski14} for time-reversal SETs. Two special cases of $\mu_a$ are
\begin{equation}
\mu_{\mathbbm{1}} =1, \quad \mu_{f}=-1,
\end{equation} 
where the latter follows from the anti-commutation relation of fermion creation operators. One of the constraints on $\mu_a$ is that if $a,b,c$ all have well-defined mirror eigenvalues ($\mu_a \neq 0$) and $N_{ab}^c\neq 0$, we must have
\begin{equation}
\mu_c = \mu_a\mu_b.
\end{equation}
Then, $\mu_{af} = -\mu_{a}$ since $\mu_f=-1$. Another constraint is that for Abelian anyons,  if $b\times \rho_m(b) = a$, we have $\xi_a =1$ and 
\begin{equation}
\mu_a = \theta_a .
\end{equation}
A proof of the latter constraint can be found in Refs.~\cite{BarkeshliPRB2019,metlitski15,chen14a,bonderson13,ZaletelPRL2015}  in the context of time-reversal SETs. The complete set of constraints is not known yet. Nevertheless,  $\{\mu_a\}$ together with $\rho_m$ seems enough to characterize mirror SETs, i.e., no other quantities are needed. At least, they are enough to determine the anomaly indicator in \eqref{eq:etam}.

Finally, we can also stack mirror-symmetric SPTs. It is known that 2D fermionic SPT phases with $\mathcal{M}^2 =1$ have a $\Z_2$ classification. (Accordingly to the crystalline equivalence principle\cite{TCEP}, the nontrivial phase corresponds to 2D $\mathcal{T}^2 =-1$ topological superconductors.) However, stacking SPTs will not affect our discussion of mirror anomalies, so we will not look into it.

In this work, we always assume that a valid $\rho_m$ is given. Generally speaking, there are obstructions to certain seemingly valid $\rho_m$. In  bosonic systems, there is the so-called $H^3$ obstruction \cite{BarkeshliPRB2019}. In fermionic cases, it may be more complicated. We assume obstructions on $\rho_m$ do not occur. We focus on the constraints on the data $\{\mu(a)\}$, and aim to derive the anomaly indicator $\eta_\mathcal{M}$ with a valid $\rho_m$.

% Generally speaking, some seemingly valid permutation $\rho_m$ may also exhibit the so-called $H^3$ anomaly \cite{SET}. However, in this work, we only consider those $\rho_m$'s that do not lead to $H^3$ anomaly.

\begin{table*}
\caption{List of some notation and convention used in this work, for readers' quick reference.}
\begin{tabular}{l|l}
\hline\hline
$\mathcal{C}$ & original fermionic topological order\\
$\bar{\mathcal{C}}$ & the complement of $\mathcal{C}$ in $\mathcal{B}$ as a set\\
$\mathcal{B}$ & $\Z_2^f$-gauged theory of $\mathcal{C}$; as sets, we have $\mathcal{B} = \mathcal{C}\oplus \bar{\mathcal{C}}$ \\
$\mathcal{B}_l,\mathcal{B}_r$ & $\Z_2^f$-gauged theories of $\mathcal{C}$ on the two sides of the mirror plane respectively; $ \mathcal{B}\equiv\mathcal{B}_l$ in most discussions \\
$\mathcal{B}\boxtimes\mathcal{B}$ & double-layer topological order obtained by folding along the mirror axis, after proper relabeling \\
$\mathcal{D}$ & topological order obtained by gauging $\Z_2^{\rm ex}$ layer-exchange symmetry in $\mathcal{B}\boxtimes\mathcal{B}$ \\
$a,b,\dots$ & anyons in $\mathcal{C}$ \\
$v_1,v_2,\dots$ & fermion parity vortices in $\mathcal{B}$; $v_i\in\bar{\mathcal{C}}$ \\
$\alpha, \beta, \dots$ & general anyons in $\mathcal{B}$, or anyons in general topological orders\\
$[a]$ & the pair $\{a, af\}$ for $a\in\mathcal{C}$\\
$[v]$ & the pair $\{v,vf\}$  for non-Majorana-type vortices; or simply $v$ itself for Majorana-type vortices\\
%$S_{\alpha,\beta}$ & elements of the $S$ matrix of $\mathcal{B}$ \\
%$\mathcal{S}(\alpha, \beta)$ & elements of the $S$ matrix of $\mathcal{D}$\\
$c$ & when referred as chiral central charge, it is the chiral central charge of $\mathcal{B}$\\
$\mu_a$ & mirror eigenvalue of certain two-anyon state, defined in Eqs.~\eqref{eq:mirroreig} and \eqref{eq:mirroreig2} \\
$\xi_a$ & defined in \eqref{eq:xi}\\
$\sigma_v$ & a quantity that denotes whether $v$ is of Majorana or non-Majorana type, defined in \eqref{eq:sigma_v}\\
$\Lambda_{a,v}$ & a block of the $S$ matrix of $\mathcal{B}$ after renormalization, defined in \eqref{eq:lambda_def}\\
$n_v,p_v,n_v',p_v'$ & restriction/lifting coefficients defined in \eqref{eq:gr-vortex}\\
$\hat{n}_v, \hat{p}_v$ & integers defined in \eqref{eq:hatn-def} and \eqref{eq:hatp-def} \\
\hline\hline
\end{tabular}
\end{table*}

\subsection{Gauging $\Z_2^f$ before folding}
\label{sec:gauging_z2f}

As discussed in Sec.~\ref{sec:dimreda}, to study mirror SETs, our strategy is to perform proper LUTs and then fold the surface along the mirror axis as in Fig.~\ref{fig:folding}. The folded double-layer topological order has a $\Z_2^f \times \Z_2^{\rm ex}$ internal symmetry, so we can further analyze it by gauging the symmetries. However, we find it more convenient to gauge the fermion parity $\Z_2^f$ before folding. Therefore, here we discuss properties of the mirror SETs after gauging $\Z_2^f$.  We will see later that gauging $\Z_2^f$ before and after folding work equally well for our purpose. 

To gauge $\Z_2^f$, we couple the system to a dynamical gauge field through the minimal coupling procedure (see Refs.~\onlinecite{levin2012, WangPRB2015} for technical details). After gauging, new excitations, namely fermion parity vortices, are introduced into the system. We denote the enlarged topological order as $\mathcal{B}$. We list a few properties of $\mathcal{B}$ below; more can be found, e.g., in Refs.~\onlinecite{Bruillard2017,LanPRB2016}. It contains  the original anyons in $\mathcal{C}$ and the new fermion-parity vortices, 
\begin{equation}
\mathcal{B}= \{\underbrace{1, f, a, af, \dots}_{\mathcal{C}},\underbrace{v_1, v_1f, \dots}_{\text{non-Majorana}}, \underbrace{v_j,v_{j+1}, \dots}_{\text{Majorana}}\}.
\end{equation}
We distinguish two types of vortices, ``Majorana type'' and ``non-Majorana type''. Non-Majorana-type vortices come in pairs $v_i$ and $v_if$, while Majorana-type vortices come alone. This distinction is characterized by the fusion rule between vortex $v$ and $f$:
\begin{equation}
v\times f = \left\{
\begin{array}{ll}
v, & \quad \text{Majorana type,}\\
vf, & \quad \text{non-Majorana type.}
\end{array}
\right.
\end{equation}
For non-Majorana vortices, we have  $\theta_v = \theta_{vf}$. The topological order $\mathcal{B}$ is \emph{modular}, i.e., it should be viewed as a bosonic topological order. The fermion $f$ is not transparent any more after gauging $\Z_2^f$. In particular, its mutual statistics with respect to any vortex is given by
\begin{equation}
M_{f,v} =-1,
\label{eq:monodromy2}
\end{equation}
which is essentially the Aharonov-Bohm phase in $\Z_2$ gauge theories. Another property is that
\begin{equation}
\sum_{a\in\mathcal{C}} d_a^2 = \sum_{v\in\bar{\mathcal{C}}} d_{v}^2,
\end{equation}
where $\bar{\mathcal{C}}$ is the complement of $\mathcal{C}$ in $\mathcal{B}$. Accordingly, the total quantum dimension $D_\mathcal{B} = \sqrt{2}D_\mathcal{C}$.

Like in $\mathcal{C}$, we use the notation $[v]$ to denote the pair $\{v, vf\}$ for non-Majorana-type vortices; for the Majorana type, $[v]$ contains only one vortex. Regardless of this, we will simply say ``$[v]$ is a pair''.  The total number of vortex pairs is $|\mathcal{C}|/2$, the same as pairs of the original anyons in $\mathcal{C}$  \cite{Bruillard2017}. Below, we will also use the notation $vf$ for Majorana-type vortices to simplify discussions. However, it should be noted that $vf$ and $v$ are the \emph{same} anyon for Majorana-type vortices.   We find it convenient to define the following quantity
\begin{equation}
\sigma_v  = \left\{
\begin{array}{ll}
1, & \quad \text{if $v$ is non-Majorana,}\\
\sqrt{2}, & \quad\text{if $v$ is Majorana.}
\end{array}
\right.
\label{eq:sigma_v}
\end{equation}
It is obvious that $\sigma_v = \sigma_{vf}$. With this, we define a matrix $\Lambda$,
\begin{equation}
\Lambda_{a,v} = \frac{2}{\sigma_v}S_{a,v} ,
\label{eq:lambda_def}
\end{equation}
where $a\in\mathcal{C}$, $v\in\bar{\mathcal{C}}$, and $S_{a,v}$ is the $S$ matrix of $\mathcal{B}$. That is, $\Lambda$ is the off diagonal part of $S$ between original anyons and fermion-parity vortices, up to renormalization. An important property of $\Lambda$ is that
\begin{align}
\sum_{[a]}\Lambda_{a,v}\Lambda^*_{a,v'}  & = \delta_{[v],[v']}, \nonumber\\
\sum_{[v]}\Lambda_{a,v}\Lambda^*_{a',v}  & = \delta_{a,a'} - \delta_{af,a'} ,
\label{eq:lambda_unitarity}
\end{align}
where the summation over $[a]$ or $[v]$ means that only one anyon in each pair is summed over. We prove this property in Appendix \ref{sec:app_B_prop}. The sum does not depend on which anyon in $[a]$ (and $[v]$) is summed over, due to the following properties
\begin{equation}
\Lambda_{a,v}  = \Lambda_{a,vf}= -\Lambda_{af,v}.
\end{equation}
One may view $\Lambda$ as a unitary matrix, if we only count $[a]$ as its index. We will assume that $\Lambda$ is given in our later discussions. In fact, one can show \cite{Wang_unpub_2020} that it is not hard to derive $\Lambda$ by analyzing the Verlinde algebra associated with $\mathcal{C}$ (see some discussions in Appendix \ref{sec:app_B_prop}).

It is worth mentioning that  $\mathcal{B}$ is not unique for a given $\mathcal{C}$ (i.e., given all the fusion and braiding data). It is proven that if there exists one\footnote{Whether there exists at least one modular extension $\mathcal{B}$ of a fermionic topological order remains an open question. It is conjectured to be true. In this work, we take the assumption that a valid $\mathcal{B}$ always exists.}, there always exists $16$ distinct $\mathcal{B}$'s\cite{Bruillard2017, LanPRB2016}.  The sixteen possible $\mathcal{B}$'s differ from each other in their chiral central charge $c$ by a multiple of $1/2$, up to modulo 8 (as well as other properties such as topological spin of vortices, etc). It is well known that $c$ modulo 8 of a UMTC can be computed from its fusion  and braiding data as follows
\begin{equation}
e^{i2\pi c/8} = \frac{1}{D_\mathcal{B}} \sum_{\alpha\in \mathcal{B}} d_\alpha^2 \theta_\alpha .
\label{eq:centralcharge}
\end{equation}
More explicitly, the sixteen possible $\mathcal{B}$'s can be obtained by gauging $\Z_2^f$ in $\mathcal{C}$ after it is stacked with multiple copies of $p_x+ip_y$ superconductors. Stacking $p_x+ip_y$ superconductors onto $\mathcal{C}$ does not change the fusion and braiding data,  but it does change the central charge $c$ of its edge theory by multiples of $1/2$. (Note that $c$ does not change after gauging $\Z_2^f$.) However, due to adjoining operations (Fig.~\ref{fig:adjoining}),  the central charge $c$ before gauging $\Z_2^f$ is fixed only up to modulo $1/2$. Accordingly, any of the sixteen $\mathcal{B}$'s is a valid $\Z_2^f$-gauged theory of $\mathcal{C}$. Nevertheless, we expect that the final result on mirror anomaly does not depend on which $\mathcal{B}$ we take, since the adjoining operation does not change the relevant physics of mirror SETs.

Now we consider mirror symmetry properties of the $\Z_2^f$-gauged theory. From Fig.~\ref{fig:folding}(b), we see that left and right parts have opposite chiralities after proper LUTs. That is, if only fusion and braiding statistics are considered, the left and right parts both host a fermionic topological order $\mathcal{C}$, but they are associated with opposite chiral central charges, $c_r = - c_l$. After gauging $\Z_2^f$, $c_l$ and $c_r$ do not change. In general, the left and right parts have different gauged topological orders, which we denote by $\mathcal{B}_l$ and $\mathcal{B}_r$ respectively. Both $\mathcal{B}_l$ and $\mathcal{B}_r$ are among the 16 possible $\Z_2^f$-gauged theories of $\mathcal{C}$. Moreover, $\mathcal{B}_r$ is determined once we fix $\mathcal{B}_l$, since $c_r=-c_l$ and chiral central charge uniquely determines the $\Z_2^f$-gauged theory of $\mathcal{C}$.  In fact, due to the mirror symmetry,  $\mathcal{B}_r$ must be the mirror image of $\mathcal{B}_l$. Accordingly, we can extend the domain of the map $\rho_m$ from $\mathcal{C}$ to $\mathcal{B}_l$. We define 
\begin{equation}
\rho_m: \mathcal{B}_l \rightarrow \mathcal{B}_r,
\label{eq:rho-ext}
\end{equation}
such that if $a\in\mathcal{C}$, we require $\rho_m(a)\in \mathcal{C}\subset \mathcal{B}_r$ to reduce to the original map $\rho_m$.\footnote{Generally speaking, for given $\mathcal{B}_l$ and $\mathcal{B}_r$, when we extend the domain from $\mathcal{C}$ to $\mathcal{B}_l$, the extended map $\rho_m$ is not unique. However, it is not hard to see distinct extensions  differ by some auto-equivalences of $\mathcal{B}_r$, which is irrelevant for our later discussions.}  The property $\theta_{\rho_m(\alpha)} = \theta_\alpha^*$ still holds, and it holds for every anyon $\alpha\in \mathcal{B}_l$ including the fermion parity vortices.  

%In fact, we will abuse notations and denote the anyons in $\mathcal{B}_r$ by $\rho_m(a)$, as the map $\rho_m$ is one-to-one.

Note that $\rho_m$ is a map between two topological orders $\mathcal{B}_l$ and $\mathcal{B}_r$. In general, $\mathcal{B}_r\neq \mathcal{B}_l$, i.e.,  after gauging $\Z_2^f$ the left and right part of the surface are not equivalent. Therefore, after the extension of its domain, $\rho_m$ is not an anti-autoequivalence any more.  Nevertheless, both $\mathcal{B}_l$ and $\mathcal{B}_r$ are obtained from the same $\mathcal{C}$, so their central charges can only differ by multiples of $1/2$. Since $c_l + c_r =0$, we must have $c_l, c_r$ be multiples of $1/4$. In Appendix \ref{sec:app_B_prop}, we also prove some properties of the $\Lambda$ matrix introduced in \eqref{eq:lambda_def}, resulting from the presence of mirror symmetry in $\mathcal{C}$.

\subsection{Folding the $\Z_2^f$-gauged theory}
\label{sec:folding-gauged}

Next we fold the $\Z_2^f$-gauged topological order. To be precise, we turn $\mathcal{B}_r$ round so that it is below $\mathcal{B}_l$ [Fig.~\ref{fig:folding}(c)]. Since folding reverses the spatial orientation, the double-layer topological order is actually  $\mathcal{B}_l \boxtimes \mathcal{B}_r^{\rm rev}$, which means a simple stack of $\mathcal{B}_l$ and $\mathcal{B}_r^{\rm rev}$. Topological spins of the anyons in $\mathcal{B}_r^{\rm rev}$ should get complex conjugated due to the reverse of orientation. More specifically, if $\alpha\in \mathcal{B}_r$, we denote its reverse by $\alpha^{\rm rev}\in \mathcal{B}_r^{\rm rev}$. Then, $\theta_{\alpha^{\rm rev}} = \theta_\alpha^*$. 

We observe that $\mathcal{B}_r^{\rm rev}$ is equivalent to $\mathcal{B}_l$. To see that, we notice that reversing the orientation of $\mathcal{B}_r$ is the same as taking its mirror image. That is, $\mathcal{B}_r^{\rm rev}$ is actually the mirror image of $\mathcal{B}_r$. At the same time, $\mathcal{B}_r$ is in turn the mirror image of $\mathcal{B}_l$. Accordingly, $\mathcal{B}_r^{\rm rev}$ is equivalent to $\mathcal{B}_l$. More specifically, under the composite action of $\rho_m$ and ``$\rm rev$'', $\alpha\in \mathcal{B}_l$ is mapped to  $[\rho_m(\alpha)]^{\rm rev}$. 
We know that $\left[\rho_m(\alpha)\right]^{\rm rev}$ is actually the double mirror image of $\alpha$. So, they are actually the same, except that $\alpha$ lives in the top layer, while $\left[\rho_m(\alpha)\right]^{\rm rev}$ lives in the bottom layer.  It is convenient to make the relabeling for anyons in $\mathcal{B}_r^{\rm rev}$
\begin{equation}
 \left[\rho_m(\alpha)\right]^{\rm rev} \leftrightarrow \alpha.
\label{eq:identification}
\end{equation}
Then, we will simply denote the double-layer topological order by $\mathcal{B}\boxtimes \mathcal{B}$, where we have suppressed the index $l$ in $\mathcal{B}_l$. Anyons in the double-layer topological order then can be denoted as $(\alpha, \beta)$, where $\alpha,\beta\in\mathcal{B}$.

After folding, the mirror symmetry $\mathcal{M}$ becomes an layer-exchange symmetry of the double-layer topological order. We denote the symmetry group as $\Z_2^{\rm ex}$. It is not hard to see that the mirror symmetry permutes anyons as follows
\begin{equation}
\mathcal{M}: \ (\alpha,\beta) \leftrightarrow (\beta,\alpha).
\label{eq:m-permute}
\end{equation}
Only the diagonal anyon $(\alpha,\alpha)$ is invariant under layer exchange. 

\begin{figure}
\centering
\includegraphics[scale=1]{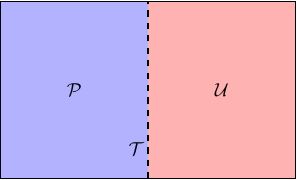}
\caption{Two topological orders $\mathcal{P}$ and $\mathcal{U}$ connected by a gapped domain wall, on which lives a fusion category $\mathcal{T}$. }
\label{fig:anyoncond}
\end{figure}

We see that $\rho_m$ does not enter the above symmetry permutation. It turns out that $\rho_m$ enters the properties of the gapped domain wall in Fig.~\ref{fig:folding}(c), as expected from our conclusion in Sec.~\ref{sec:dimreda} that all SET information is encoded at the gapped domain wall. One way to describe a gapped domain wall is to use the theory of \emph{anyon condensation} \cite{bais2009}. Qualitatively, it means that a subset of anyons can be ``condensed'' when they move to the domain wall, causing a ``quantum phase transition'' to a new topological order living on the other side of the domain wall (see a sketch in Fig.~\ref{fig:anyoncond}; certain anyons in $\mathcal{P}$ are condensed, giving rise to $\mathcal{U}$). That is, we can use local operators at the domain wall to destroy (and create) the condensed anyons. They become local excitations, i.e., the new vacuum anyon. At the same time, other anyons may be identified with and/or split into the anyons that live on the other side. Anyon condensation theory is such a theory that describes the relation between the topological orders on the two sides of the domain all. In Sec.~\ref{sec:anyoncon}, we will study in detail about anyon condensation. Below,  we only give a qualitative description and show how the SET quantities, $\rho_m$ and $\{\mu_a\}$, enter the properties of the gapped domain wall.

Let us consider the double-layer topological order $\mathcal{C}\boxtimes\mathcal{C}$ before gauging. We claim that the condensed anyons are of the form $(a, \bar\rho_m(a))$, for every $a\in\mathcal{C}$. This can be seen as follows: Consider two anyons, $a$ on the left side and $b$ on the right side of the mirror axis, before folding. When they move close and meet at the mirror axis, they can annihilate together and fuse into the vacuum channel if and only if $b=\bar{a}$. After folding and the relabelling \eqref{eq:identification}, this condition translates into that an anyon $(a, b)$ can be annihilated (condensed) at the domain wall if and only if $b=\bar\rho_m(a)$. We can write the condensate in a more concise form through the so-called \emph{lifting} map
\begin{equation}
l({1}) = \sum_{a\in\mathcal{C}} (a, \bar \rho_m(a)),
\label{eq:lifting1}
\end{equation}
where  ``${1}$'' denotes the new vacuum after condensation.  At the same time, before folding, $a$ and $b$  can fuse together into the $f$ channel, if and only if $b = \bar{a}f$. So, in the folded picture, we should have
\begin{equation}
l(f) = \sum_{a\in\mathcal{C}} (af, \bar \rho_m(a)).
\label{eq:lifting2}
\end{equation}
That is, anyons in $l(f)$ can be identified as $f$ when they move to domain wall, upon action of proper local operators. Since $\mathcal{C}\boxtimes \mathcal{C}$ is a subset of anyons in $\mathcal{B}\boxtimes\mathcal{B}$, which is closed under fusion and braiding, the above description on the gapped domain wall still holds even after we gauge $\Z_2^f$ in $\mathcal{C}$. In particular, other anyons in $\mathcal{B}\boxtimes\mathcal{B}$, those involving fermion-parity fluxes, will not enter $l(1)$ and $l(f)$. We leave detailed discussions in Sec.~\ref{sec:anyoncon}. 

The lifting maps \eqref{eq:lifting1} and \eqref{eq:lifting2} only encode the information of $\rho_m$. How about the information of mirror eigenvalues $\{\mu_a\}$? Recall the definition of $\mu_a$  through the two-anyon state $|a,\rho_m(a)\rangle$. Since $\xi_a$ must be nonzero in order to have a well-defined $\mu_a$ with value $\pm1$,   $a$ and $\rho_m(a)$ can only fuse into either $\mathbbm{1}$ or $f$ when they meet at the mirror axis. The obtained local excitation ($\mathbbm{1}$ or $f$) carries a mirror eigenvalue $\mu_a$. It means that, in the folded system  $\mathcal{C}\boxtimes\mathcal{C}$ after relabelling \eqref{eq:identification},  when the anyon $(a, a)$ moves to the boundary, it ``condenses'' into a local excitation (either $\mathbbm{1}$ or $f$) that carries a $\Z_2^{\rm ex}$ charge $\mu_a$. To have a symmetric gapped domain wall, only neutral anyons can be condensed. So,  $(a,a)$ should be condensed together with an extra $\Z_2^{\rm}$ charge $\mu_a$. We see that the $\Z_2^{\rm ex}$ symmetry requirement of the gapped domain wall encodes the information $\{\mu_a\}$. Later we will use the notation $(a,a)^{\pm}$, where the sign ``$\pm$'' denotes if $(a,a)$ carries a $\Z_2^{\rm ex}$ charge or not. Then, at the domain wall, the condensed anyons should be $(a,a)^{\mu_a}$ if $\xi_a\neq0$.  To better deal with the symmetry issue, we will further gauge $\Z_2^{\rm ex}$ in the next subsection.

Before proceeding, we find it a good place to spend some words discussing why gauging $\Z_2^f$ before folding and after folding work equally well. If the surface is folded before gauging $\Z_2^f$, we obtain a simple stack  $\mathcal{C}\boxtimes\mathcal{C}$ after relabelling \eqref{eq:identification}. Each layer in the stack has a fermion-parity conservation symmetry, which we denote as $\Z_2^{\rm top}$ and $\Z_2^{\rm bot}$, respectively. More explicitly, the fermion parity operators are $(-1)^{N_{\rm top}}$ and $(-1)^{N_{\rm bot}}$, where $N_{\rm top}$ and $N_{\rm bot}$ are the fermion numbers in the two layers. By isomorphism, we can equivalently write the symmetry group  as $\Z_2^{f}\times\Z_2^{\rm extra}$, where we take $(-1)^{N_{\rm top}+ N_{\rm bot}}$ as the non-trivial element in $\Z_2^f$, and $(-1)^{N_{\rm top}}$ as the non-trivial element in $\Z_2^{\rm extra}$. That is,  $\Z_2^{\rm extra} = \Z_2^{\rm top}$. One can easily observe that the double-layer topological order $\mathcal{B}\boxtimes\mathcal{B}$ is the theory obtained from $\mathcal{C}\boxtimes\mathcal{C}$ by gauging $\Z_2^{f}\times\Z_2^{\rm extra}$. That is, the extra $\Z_2^{\rm extra}$ symmetry is also gauged. In $\mathcal{B}\boxtimes\mathcal{B}$, the $\Z_2^{\rm extra}$ charge is the boson $(f,f)$, while the $\Z_2^f$ charge is $(1,f)$.  To restore a theory with only $\Z_2^f$ gauged, one can condense $(f,f)$ in $\mathcal{B}\boxtimes\mathcal{B}$. Nevertheless, we have already included it in $l(1)$ in \eqref{eq:lifting1}. Therefore, condensing anyons in $l(1)$ will actually restore a $\Z_2^f$-gauged theory, leading to our conclusion that gauging $\Z_2^f$ before and after folding is equivalent. In this argument, we have implicitly used the fact that doing anyon condensation in a whole or in parts with different orders,  gauging symmetries in a whole or in parts with different orders, and doing anyon condensation and gauging interchangeably, all commute. In fact, below we will further gauge $\Z_2^{\rm ex}$, and consider anyon condensation after gauging $\Z_2^{\rm ex}$ in Sec.~\ref{sec:z2^ex}.

\subsection{Further gauging $\Z_2^{\rm ex}$ symmetry}
\label{sec:D}

Above, we have obtained a double-layer topological order $\mathcal{B}\boxtimes\mathcal{B}$ by folding. It has a  $\Z_2^{\rm ex}$ layer-exchange symmetry that permutes the anyons according to \eqref{eq:m-permute}. The gapped domain wall in Fig.~\ref{fig:folding}(c) is also symmetric and encodes the information of $\rho_m$ and $\{\mu_a\}$.  To better analyze the symmetric gapped domain wall, we follow Ref.~\onlinecite{folding} and gauge the $\Z_2^{\rm ex}$ symmetry. We are able to do so because folding has turned $\mathcal{M}$ into an internal symmetry. We realize that gauging $\Z_2^{\rm ex}$ in $\mathcal{B}\boxtimes\mathcal{B}$ is exactly the same as in the bosonic case. The $\Z_2^{\rm ex}$-gauged theory of $\mathcal{B}\boxtimes\mathcal{B}$ has been explicitly obtained in Ref.~\onlinecite{folding}. So, we just need to borrow the result from there. This is also one of the reasons that we gauge $\Z_2^f$ before folding, as it simplifies our computation. Below we review a few properties of the $\Z_2^{\rm ex}$-gauged theory, which we denote as $\mathcal{D}$.

The $\Z_2^{\rm ex}$-gauged theory $\mathcal{D}$  consists of anyons that originate from $\mathcal{B}\boxtimes\mathcal{B}$, as well as new anyons which are vortices of the $\Z_2^{\rm ex}$ symmetry. We list them below. First, for each pair $\alpha, \beta\in\mathcal{B}$, with $\alpha\neq \beta$, we have an anyon $[\alpha,\beta]$ in $\mathcal{D}$. This anyon originates from the symmetrization of $(\alpha,\beta)$ and $(\beta, \alpha)$ from $\mathcal{B}\boxtimes\mathcal{B}$.  It has a quantum dimension $d_{[\alpha,\beta]} = 2d_\alpha d_\beta$ and a topological spin $\theta_{[\alpha,\beta]} = \theta_\alpha\theta_\beta$. Second, each diagonal anyon $(\alpha,\alpha)\in\mathcal{B}\boxtimes\mathcal{B}$ becomes $(\alpha,\alpha)^{\pm}$, where $\pm$ represents the $\Z_2^{\rm ex}$ charge that it carries. The quantum dimension is $d_{(\alpha,\alpha)^\pm} = d_\alpha^2$ and topological spin is $\theta_{(\alpha,\alpha)^\pm} = \theta_\alpha^2$. Third, there are $2|\mathcal{B}|$ distinct $\Z_2^{\rm ex}$ vortices, which we denote as $X_\alpha^\pm$, $\forall \alpha\in\mathcal{B}$. The quantum dimension is $d_{X_\alpha^\pm} = d_\alpha D_\mathcal{B}$, and the topological spin is given by
\begin{equation}
\theta_{X_\alpha^\pm} = \pm e^{i2\pi c/16} \sqrt{\theta_\alpha},
\label{eq:vortexspin}
\end{equation}
where $c$ is the chiral central charge of $\mathcal{B}$. In total, there are $|\mathcal{B}|(|\mathcal{B}|+7)/2$ anyons. One can check that the total quantum dimension $D_\mathcal{D} = 2D_\mathcal{B}^2$. We comment that ``$\pm$'' on $X_\alpha^\pm$ is conventional and has no absolute meaning. Relatively, $X_\alpha^-$ differs from $X_\alpha^+$ by a $\Z_2^{\rm ex}$ charge, which is the Abelian anyon $(\mathbbm{1}, \mathbbm{1})^-$. On the other hand, the ``$\pm$'' sign on $(\alpha,\alpha)^\pm$ does have an absolute meaning of $\Z_2^{\rm ex}$ charge, at least for the anyons satisfying $\bar\rho_m(a) = a$ or $a f$, with $a\in\mathcal{C}$. For these anyons, the $\Z_2^{\rm ex}$ charge inherits from the mirror eigenvalue through folding.

Several fusion and braiding properties are in order. A few useful fusion rules are
\begin{align}
(\alpha,\alpha)^{\pm}\times (\mathbbm{1}, \mathbbm{1})^- &  = (\alpha,\alpha)^{\mp},\nonumber\\
[\alpha,\beta]\times (\mathbbm{1}, \mathbbm{1})^- &  = [\alpha,\beta],\nonumber\\
X_{\alpha}^\pm \times (\mathbbm{1}, \mathbbm{1})^- &  =  X_\alpha^{\mp},\nonumber\\
(\alpha,\alpha)^{\pm}\times (f, f)^+ &  = (\alpha f,\alpha f)^{\pm},\nonumber\\
[\alpha,\beta]\times (f, f)^+ &  = [\alpha f,\beta f],\nonumber\\
X_{a}^\pm \times (f, f)^+&  =  X_{a}^{\pm},\nonumber\\
X_{v}^\pm \times (f, f)^+&  =  X_{v}^{\mp},\nonumber\\
%(\alpha,\alpha)^{\pm}\times [1, f] &  = \left\{
%\begin{array}{ll}
%(\alpha,\alpha)^+ + (\alpha,\alpha)^- & \text{if }\alpha =\alpha f \\[3pt]
%[\alpha, \alpha f]
%\end{array}
%\right.\nonumber\\
%[\alpha,\beta]\times [1, f] &  = [\alpha f,\beta] + [\alpha, \beta f]\nonumber\\
X_{\alpha}^\pm \times [1, f] &  =  X_{\alpha f}^{+} + X_{\alpha f}^-,
\end{align}
where $\alpha,\beta$ are general anyons in $\mathcal{B}$, $a\in\mathcal{C}$ and $v\in\bar{\mathcal{C}}$. In the case that $\alpha$ is a Majorana-type fermion-parity vortex, $\alpha f\equiv\alpha$ is understood. The $T$ matrix is determined by the topological spins given above. The $S$ matrix of $\mathcal{D}$ can be expressed in terms of that of $\mathcal{B}$. In particular, we have
\begin{align}
S^\mathcal{D}(X_\alpha^\pm, [\beta,\gamma]) & = 0 ,\nonumber\\
S^\mathcal{D}(X_\alpha^\pm, (\beta,\beta)^\mu) & =  \frac{\mu}{2} S^\mathcal{B}_{\alpha, \beta},
\label{eq:smatrix-of-D}
\end{align}
where $\mu = \pm$, and we use $S^{\mathcal{D}}(X, Y)$ instead of $S^\mathcal{D}_{X,Y}$ for notational clarity.  Readers can find more properties of $\mathcal{D}$ in Ref.~\onlinecite{folding}.

\section{Anyon condensation at the domain wall}
\label{sec:anyoncon}

In this section, we study properties of the domain wall in detail through the theory of anyon condensation. These properties, briefly discussed in Sec.~\ref{sec:folding-gauged}, serve as boundary conditions of the double-layer topological order, and will be used to determine the iTO on the right-hand side of Fig.~\ref{fig:folding}(c). 

\subsection{Review on anyon condensation}

We briefly review the theory of anyon condensation here. For more details, one may consult Refs.~\cite{bais2009,EliensPRB2014, Kong2014, lan2015,HungJHEP2015, NeupertPRB2016}.

Anyon condensation is a quantum phase transition between two topologically ordered phases. Similarly to the usual Bose-Einstein condensation, it occurs when a set of bosons condense, such that the system transits from one topological order to another topological order. Anyon condensation is most commonly studied in the scenario in Fig.~\ref{fig:anyoncond}: condensation has occurred in one part of the system (red region), but not in the other (blue region), giving rise to a domain wall between them. It considers the case that the domain wall is energetically gapped. This system can be described by three tensor categories \cite{bais2009}:
\begin{eqnarray}
\mathcal{P}\rightarrow\mathcal{T}\rightarrow\mathcal{U},
\label{eq:PTU}
\end{eqnarray}
where $\mathcal{P}$ describes the ``parent'' topological order before condensation, $\mathcal{U}$ describes the ``child'' topological order after condensation, and $\mathcal{T}$ describes the domain wall. Both $\mathcal{P}$ and $\mathcal{U}$ are UMTCs with well-defined fusion and braiding properties of anyons\cite{kitaev2006}. On the other hand, $\mathcal{T}$ is a unitary fusion category, and not all anyons in $\mathcal{T}$ are associated with braiding properties.

%$\mathcal{P}$ is a unitary modular tensor category (UMTC)in which the anyons have well defined fusion and braiding rules. After anyon condensation, we should relabel these anyons and they can form a new unitary fusion category(UFC) $\mathcal{T}$.
%Using whether the anyons have well-defined braiding structure as a criteria, we can separates the anyons in $\mathcal{T}$ into two sectors. The sector satisfing the criteria corresponds to a new UMTC $\mathcal{U}$. 

Let us understand the relations between the three categories in \eqref{eq:PTU}. When an anyon $\alpha$ in $\mathcal{P}$ moves to the domain wall, it will be identified as and/or split to some anyons in $\mathcal{T}$.  This identification/splitting is described by the \textit{restriction} map $r:\mathcal{P}\rightarrow\mathcal{T}$, defined as 
\begin{equation}
r(\alpha)=\sum_{t\in\mathcal{T}}n_{\alpha,t}t,
\label{eq:r_map}
\end{equation}
where $n_{\alpha, t}$ is an integer coefficient. At the same time, there is a reverse map called the \textit{lifting} map, which reads
\begin{equation}
l(t)=\sum_{\alpha\in\mathcal{P}}n_{\alpha,t}\alpha.
\label{eq:l_map}
\end{equation}
We will use the notation that $\alpha\in l(t)$ if $n_{\alpha,t}\neq 0$. From the two maps, we see that any anyon in $\mathcal{P}$ can be viewed as a linear superposition of anyons in $\mathcal{T}$, and vice versa. In particular, every $\alpha$ with $n_{\alpha,{1}}\neq0$ can become the vacuum anyon $1$ in $\mathcal{T}$. That is, all anyons in $l(1)$ are ``condensed''. It is required that $\mathbbm{1}\in l(1)$, where $\mathbbm{1}$ is the vacuum anyon of $\mathcal{P}$. Next, we ask whether anyons in $\mathcal{T}$ can exit the domain wall and move freely into the region of $\mathcal{U}$? If so, it is said that these anyons are \emph{deconfined}; if not, they are \emph{confined}. Deconfined anyons form the category $\mathcal{U}\subset\mathcal{T}$.  The criterion to determine whether an anyon in $\mathcal{T}$ is confined or not is as follows: $t$ is deconfined if and only if all $\alpha\in l(t)$ have the same topological spin. Then, $\theta_t = \theta_\alpha$ for any $\alpha\in l(t)$ if $t$ is deconfined. It is required that the vacuum anyon $1$ must be deconfined. Since $\mathbbm{1}\in l(1)$, we see that all anyons in $l(1)$ must be bosons. If there is an Abelian boson $b\in l(1)$, one can show that $\alpha$ is confined if it has non-trivial mutual statistics with respect to $b$. 

%These anyons included in category $\mathcal{U}$ are called \textit{deconfined} while other anyons left are called \textit{confined}. Thus the key point for studying the relation between these categories is to encode the information connecting $\mathcal{P}$ and $\mathcal{T}$.

We see that the key quantity to describe anyon condensation is $n_{\alpha,t}$, a $|\mathcal{P}|\times|\mathcal{T}|$ non-negative integer matrix\footnote{It is known that $n_{\alpha,t}$ does not provide a complete description of anyon condensation. However, for our purpose, it is enough for the derivation of mirror anomaly.}. Here we list two of its constraints. The first constraint is that  fusion commutes with the restriction map 
\begin{equation}
r(\alpha)\times r(\beta) = r(\alpha\times \beta).
\label{eq:rr=r}
\end{equation}
The explicit expression is 
\begin{equation}
\sum_{r,s\in \mathcal{T}}n_{\alpha,r}n_{\beta,s}N^{t}_{rs}=\sum_{\gamma\in \mathcal{P}}N^{\gamma}_{\alpha\beta}n_{\gamma,t},
\label{eq:fusion_PT}
\end{equation}
where $N^{t}_{rs}$ and $N^{\gamma}_{\alpha\beta}$ are the fusion coefficients in $\mathcal{P}$ and $\mathcal{T}$, respectively. Since $\mathcal{U}$ is a subcategory of $\mathcal{T}$,    Eq.~\eqref{eq:fusion_PT}  still holds if we restrict $r,s, t$ to be inside $\mathcal{U}$. Another constraint is that when restricted to $\mathcal{U}$, the matrix $n$ commutes with the modular $S$ and $T$ matrices, in the following sense 
\begin{equation}
S^{\mathcal{P}}n=nS^{\mathcal{U}},\quad T^{\mathcal{P}}n=nT^{\mathcal{U}}.
\end{equation}
The explicit expression for $S$ is
\begin{equation}
\sum_{\beta\in \mathcal{P}} S_{\alpha, \beta} n_{\beta, t} = \sum_{s\in\mathcal{U}} n_{\alpha,s} S_{s,t},
\label{eq:sn=ns}
\end{equation}
where $\alpha\in\mathcal{P}$ and $t\in\mathcal{U}$. Since $T_{\alpha,\beta}=\delta_{\alpha,\beta}\theta_{\alpha}$, the above equation for $T^\mathcal{P}$ and $T^{\mathcal{U}}$ means that $\theta_{\alpha} =\theta_t$ as long as $n_{\alpha,t}\neq 0$, for any $\alpha$ and $t$. In particular, all anyons in $l(t)$ have the same topological spin, if $t$ is deconfined.

\subsection{Ignoring $\Z_2^{\rm ex}$}
\label{sec:ignorez2ex}

%\begin{table}
%\caption{Anyons in $\mathcal{U}$ when only $\Z_2^f$ is gauged and $\Z_2^{\rm ex}$ is ignored.}
%\begin{tabular}{ll}
%\hline\hline
%$c^\mathcal{U}$ being half-integer, $\mathcal{U}= \{1, f, w\}$\\
%$1\times f = f$,  $1\times w = w$
%\end{tabular}
%\end{table}

To begin, we apply the anyon condensation theory to a simpler situation, where we ignore $\Z_2^{\rm ex}$ symmetry in the folded system. That is, we consider the setup in Fig.~\ref{fig:anyoncond} with $\mathcal{P} = \mathcal{B}\boxtimes\mathcal{B}$ and $\mathcal{U}$ being a $\Z_2^f$-gauged iTO. While 2D fermionic iTOs have a $\Z$ classification,  gauging $\Z_2^f$ results in only 16 possible topological orders\cite{kitaev2006}. The topological order $\mathcal{U}$ contains three anyons $\{1, f, w\}$ if it has a half-integer chiral central charge, or four anyons $\{1, f, w, wf\}$ if it has an integer chiral central charge. In the former case, $w$ is a Majorana-type vortex with quantum dimension $d_w=\sqrt{2}$. In the latter case, $w$ and $wf$ are two non-Majorana-type vortices with quantum dimension $d_w=d_{wf} = 1$. The sixteen gauged theories can be distinguished by the topological spin of $w$: 
\begin{equation}
\theta_{w} = e^{i 2\pi c^{\mathcal{U}}/8},
\end{equation}
where $c^{\mathcal{U}}$ denotes the chiral central charge associated with $\mathcal{U}$. For non-Majorana-type vortices, one can check that $\theta_{wf} = \theta_{w}$.  

Given $\mathcal{P} = \mathcal{B}\boxtimes\mathcal{B}$ and certain properties of the gapped domain wall,  our goal is to determine $\mathcal{U}$ out of its sixteen possibilities. In the current case that $\Z_2^{\rm ex}$ is ignored, there exists an easy way to accomplish this. Since $\mathcal{B}\boxtimes\mathcal{B}$ and $\mathcal{U}$ are connected by a gapped domain wall, the chiral central charges must be equal, i.e., $c^\mathcal{U} = 2c$, where $c$ is the central charge of $\mathcal{B}$. Accordingly, we must have
\begin{equation}
\theta_{w} = e^{i 2\pi c/4}.
\label{eq:parityspin}
\end{equation}
We note that the left-hand side of Eq.~\eqref{eq:parityspin} is a quantity of $\mathcal{U}$, and the right-hand side is a  quantity of $\mathcal{B}\boxtimes\mathcal{B}$. Therefore, a single quantity $c$ of $\mathcal{B}\boxtimes \mathcal{B}$ uniquely determines $\mathcal{U}$ out of its sixteen possibilities. 

Below we would like to re-derive \eqref{eq:parityspin} using anyon condensation theory. Through this exercise, we get familiar with some properties of the gapped domain wall, paving a way for the main study in the next subsection where $\Z_2^{\rm ex}$ is included (Sec.~\ref{sec:z2^ex}).  The fusion category $\mathcal{T}$ that lives on the domain wall contains anyons from the original fermion topological order $\mathcal{C}$, fermion-parity vortices $w_1,w_2,\dots$, and other anyons that carry $\Z_2^{\rm extra}$ fluxes (see the discussion at the end of Sec.~\ref{sec:folding-gauged}). We formally write this as
\begin{equation}
\mathcal{T} = \mathcal{C}\oplus \{w_1, w_2, \dots,\}\oplus\{\text{others}\}.
\nonumber
\end{equation}
It follows from the understanding that before gauging symmetries, how anyons in $\mathcal{C}\boxtimes\mathcal{C}$ restrict themselves on the domain wall is the same as how anyons in $\mathcal{C}$ fuse on the mirror axis. That is, $\mathcal{T} = \mathcal{C}$ with the braiding information in $\mathcal{C}$ omitted. Accordingly, we expect $\mathcal{T}$ to be some extended category of $\mathcal{C}$ after gauging symmetries. Since we gauge fermion parities in both layers, $\mathcal{T}$ gains additional fermion-parity vortices, as well as $\Z_2^{\rm extra}$ vortices which we do not care and list them above as ``others''.  Inside $\mathcal{T}$, only $1$, $f$ and the fermion-parity vortex $w$  (and $wf$ if non-Majorana) are deconfined. All others are confined. We will use the convention that $w_1$ is the deconfined one, i.e., $w_1 \equiv w\in \mathcal{U}$.

%It should be preserved under restriction maps, as monodromy can be defined between deconfined and confined anyons in $\mathcal{T}$ (if careful).  Accordingly, $(v_1, v_2)$ should be restricted to anyons in $\mathcal{T}$ that have $-1$ monodromy with respect to $f\in\mathcal{U}$. All anyons in $\mathcal{C}\subset\mathcal{T}$ have trivial monodromy respect to $f$. Hence, no anyons in $\mathcal{C}$ can occur in $r\{(v_1,v_2)\}$. 

We now consider the restriction maps. We claim that
\begin{subequations}
\label{eq:r}
\begin{align}
r\{(a,b)\} & = \sum_{c\in\mathcal{C}} N_{a\rho_m(b)}^c c, \label{eq:r1}\\
r\{(a,v)\} & = \text{confined anyons only},\label{eq:r2}\\
r\{(v,a)\} & = \text{confined anyons only},\label{eq:r3}\\
r\{(v_1,v_2)\} & = \sum_{w_i} n_{(v_1,v_2),w_i} w_i, \label{eq:r4}
\end{align}
\end{subequations}
where we have used $r\{\cdot\}$ to denote restriction maps for notational clarity, $N_{a\rho_m(b)}^c$ is the fusion multiplicity in $\mathcal{C}$ and $n_{(v_1,v_2),w_i}$ is some unknown integer. The restriction map \eqref{eq:r1} can be obtained by comparing with the fusion rules in $\mathcal{C}$ before folding and applying the relabelling \eqref{eq:identification}.  Special cases are 
\begin{align}
r\{(f,f)\} & = r\{(\mathbbm{1},\mathbbm{1})\} = 1,\nonumber\\
r\{(\mathbbm{1},f)\} & = r\{(f, \mathbbm{1})\}= f. 
\end{align}
To see \eqref{eq:r2} and \eqref{eq:r3}, we consider the mutual statistics
\begin{equation}
M_{(f,f),(a,v)} = M_{f,a}M_{f,v} = -1,
\label{eq:mono}
\end{equation}
which holds for arbitrary $a$ and $v$. Since $(f,f)$ is restricted to the vacuum, the nontrivial mutual statistics \eqref{eq:mono} implies that all anyons in $r\{(a,v)\}$ and $r\{(v,a)\}$ are confined. The restriction \eqref{eq:r4} is not known generally. In \eqref{eq:r4}, we have implicitly taken only $w_i$ in $r\{(v_1,v_2)\}$. It is because that $(v_1,v_2)$ is a fermion-parity vortex before condensation. It can be seen from the mutual statistics
\begin{equation}
M_{(\mathbbm{1},f),(v_1,v_2)} = -1, \quad M_{(f,f),(v_1,v_2)} =1.
\end{equation}  
Recall that $(f,f)$ is the $\Z_2^{\rm extra}$ charge and $(1,f)$ is the $\Z_2^f$ charge. From the commutativity Eq.~\eqref{eq:rr=r} between restriction and fusion, we obtain some constraints 
\begin{align}
r\{(v_1,v_2)\}  & = r\{(v_1,v_2)\} \times r\{(f,f)\}\nonumber\\
&   = r\{(v_1f,v_2f)\}, \nonumber
\end{align}
and 
\begin{align}
r\{(v_1,v_2)\} \times f & = r\{(v_1,v_2)\} \times r\{(1,f)\} \nonumber\\
 &  =  r\{(v_1,v_2f)\},\nonumber
\end{align}
where $v_if = v_i$ if $v_i$ is a Majorana-type vortex.  From these two constraints, we immediately have
\begin{align}
n_{(v_1, v_2), w}\! = \! n_{(v_1f, v_2f), w} \!=\! n_{(v_1, v_2f), wf}\! =\! n_{(v_1f, v_2), wf}, \nonumber\\
n_{(v_1, v_2), wf} \!=\! n_{(v_1f, v_2f), wf} \!=\! n_{(v_1, v_2f), w}\! =\! n_{(v_1f, v_2), w}. \label{eq:liftconstraint}
\end{align}
Depending on whether $v_1,v_2, w$ are Majorana-type or non-Majorana-type, these equations may be further reduced and related.

From the restriction maps, we can read out the lifting maps of the deconfined anyons in $\mathcal{U}$ as follows:
\begin{subequations}
\begin{align}
l(1) & = \sum_{a\in\mathcal{C}} (a, \bar{\rho}_m(a)), \\
l(f) & = \sum_{a\in\mathcal{C}} (af, \bar{\rho}_m(a)), \\
l(w) & = \sum_{v_1, v_2}  n_{(v_1,v_2), w} (v_1, v_2).
\label{eq:lifting-1f-noZ2}
\end{align}
\end{subequations}
This agrees with the expectation in Sec.~\ref{sec:folding-gauged}. When $w$ is of non-Majorana type, we also need to consider the lifting map $l(wf)$
\begin{equation}
l(wf)  = \sum_{v_1, v_2}  n_{(v_1,v_2), wf} (v_1, v_2).
\label{eq:lifting-1f-noZ2-wf}
\end{equation}
Soon we will see that the lifting coefficients $\{n_{(v_1,v_2), w}\}$ and $\{n_{(v_1,v_2), wf}\}$ are closely related.

Next, we make use of Eq.~\eqref{eq:sn=ns} to further constrain the lifting coefficients in $l(w)$ and $l(wf)$. Let $\alpha = (v_1, v_2)$ and $t = 1$ in \eqref{eq:sn=ns}. We immediately obtain 
\begin{equation}
\sum_{a\in\mathcal{C}}S[(v_1, v_2), (a, \bar\rho_m(a))] = \hat n_{(v_1, v_2)}\frac{d_w}{2}
\end{equation}
where we have used $S_{s,{1}} = d_{s}/2$ and defined
\begin{equation}
\hat{n}_{(v_1,v_2)} = \left\{
\begin{array}{ll}
n_{(v_1, v_2), w}, & \text{if $d_w=\sqrt{2}$} \\
n_{(v_1, v_2), w} + n_{(v_1, v_2), wf},  & \text{if $d_w=1$}
\end{array}
\right. \nonumber
\end{equation}
Using the relation $S[(v_1, v_2), (a, \bar\rho_m(a))]= S_{v_1, a} S_{v_2, \bar{\rho}_m(a)}$, Eq.~\eqref{eq:lambda_def}, and
Eq.~\eqref{eq:app_lambda6} from Appendix \ref{sec:app_B_prop},  we  obtain the following  key equation:
\begin{equation}
\sigma_{v_1}\sigma_{v_2}\delta_{[v_1],\rho_m([\bar{v}_2])} = \hat{n}_{(v_1, v_2)} d_w,
\end{equation}
where the permutation $\rho_m([\bar{v}_2])$ is defined in Appendix \ref{sec:app_B_prop}. We see that $\hat n_{(v_1,v_2)}$ is non-vanishing if and only if $[v_1] = \rho_m([\bar{v}_2])$. Since $\hat{n}_{(v_1,v_2)}$ must be an integer, we must have 
\begin{equation}
\frac{\sigma_{v_1}\sigma_{v_2}}{d_{w}} = \text{integer},
\end{equation}
if $\hat{n}_{(v_1,v_2)}\neq0$. That means, among $v_1,v_2$ and $w$, either none or two are of Majorana type. Accordingly, for $(v_1,v_2)$ with $\hat n_{(v_1,v_2)}\neq 0$,  we have: 
\begin{enumerate}
\item When $\sigma_{v_1}=1, \sigma_{v_2}=\sqrt{2}$, we must have $d_w = \sqrt{2}$ and $\hat n_{(v_1,v_2)}=1$. With the constraint \eqref{eq:liftconstraint}, we have $n_{(v_1,v_2),w} = n_{(v_1f,v_2),w} = 1$. It is similar for the case that  $\sigma_{v_1}=\sqrt{2}$ and $\sigma_{v_2}=1$. 

\item When $\sigma_{v_1}= \sigma_{v_2}=\sqrt{2}$, we must have $d_w = 1$ and $\hat n_{(v_1,v_2)}=2$. With the constraint \eqref{eq:liftconstraint}, we have $n_{(v_1,v_2),w} = n_{(v_1,v_2),wf} = 1$. 

\item When $\sigma_{v_1}= \sigma_{v_2}=1$, we must have $d_w = 1$ and $\hat n_{(v_1,v_2)}=1$. Without lose of generality, we can set $n_{(v_1,v_2),w} = 1$,  $n_{(v_1,v_2),wf} = 0$. Other lifting coefficients can be obtained from the constraint \eqref{eq:liftconstraint}.
\end{enumerate}
For other vortices $(v_1,v_2)$ with $[v_1] \neq \rho_m([\bar{v}_2])$, the  lifting coefficients are zero. Hence, all the lifting coefficients in \eqref{eq:lifting-1f-noZ2} and \eqref{eq:lifting-1f-noZ2-wf} are obtained.

Finally we prove Eq.~\eqref{eq:parityspin} based on the above understanding of the gapped domain wall. The topological spin of the deconfined anyon $w$ is given by 
\begin{equation}
\theta_{w} = \theta_{(v_1,v_2)} =  \theta_{v_1}\theta_{v_2},
\label{eq:thetaw}
\end{equation}
where $(v_1,v_2)\in l(w)$, i.e., $n_{(v_1,v_2),w} \neq 0$. That is, we must have $[v_1]=\rho_m([\bar v_2])$. To relate $\theta_w$ to the central charge $c$, we make use a general property of 
UMTC\cite{kitaev2006}:
\begin{equation}
e^{i2\pi c/8}d_\beta \theta_{\beta}^* = \sum_{\alpha\in\mathcal{B}} d_\alpha \theta_\alpha S_{\alpha,\beta}.
\end{equation}
Taking $\beta$ to be a vortex, we have
\begin{align}
e^{i2\pi c/8}d_v \theta_{v}^* & = \sum_{a\in\mathcal{C}} d_a \theta_a S_{a,v} + \sum_{v'\in\bar{\mathcal{C}}} d_{v'} \theta_{v'} S_{v',v} ,\nonumber\\
e^{i2\pi c/8}d_{vf} \theta_{vf}^* & = \sum_{a\in\mathcal{C}} d_a \theta_a S_{a,v} - \sum_{v'\in\bar{\mathcal{C}}} d_{v'} \theta_{v'} S_{v',v}, \nonumber
\end{align}
where have applied Eqs.~\eqref{eq:app_lambda_s} from Appendix \ref{sec:app_B_prop}. Adding up the two equations and using $d_{vf}=d_v$ and $\theta_v=\theta_{vf}$, we have
\begin{equation}
e^{i2\pi c/8 } d_{v}\theta_{v}^*  = \sum_{a\in\mathcal{C}} d_a \theta_a S_{a, v}.\label{eq:iii}
\end{equation}
Now we set $v = v_2$. Then,
\begin{align}
e^{i2\pi c/8 } d_{v_2}\theta_{v_2}^* 
& = \sum_{a\in\mathcal{C}} d_a \theta_a S_{a, v_2} \nonumber\\
& = \sum_{a\in \mathcal{C} }d_{\bar\rho_m(a)}\theta_{\bar\rho_m(a)} S_{\bar\rho_m(a), v_2}\nonumber\\
& = \sum_{a\in \mathcal{C} }d_{a}\theta_{a}^* S_{a, v_1}^* \frac{\sigma_{v_2}}{\sigma_{v_1}}\nonumber\\
& = e^{-i2\pi c/8} d_{v_1}\theta_{v_1}\frac{\sigma_{v_2}}{\sigma_{v_1}}.
\end{align}
From the second to third line, we have used the definition of $\rho_m([\bar{v}_2])$ and the requirement $[v_1] = \rho_m([\bar{v}_2])$. To obtain the last line, we have used \eqref{eq:iii} again.  Combining this with Eq.~\eqref{eq:app_lambda10} from Appendix \ref{sec:app_B_prop}, we derive
\begin{equation}
\theta_{v_1} = \theta_{v_2}^* e^{i2\pi c/4}.\nonumber
\label{eq:spinw}
\end{equation}
Further combining this equation with \eqref{eq:thetaw},  we immediately obtain \eqref{eq:parityspin}.

\subsection{With $\Z_2^{\rm ex}$}
\label{sec:z2^ex}

\begin{table}
  \caption{Quantum dimensions of vortices in $\mathbb{Z}_2^f\times \mathbb{Z}_2$-gauged iTOs. }
  \label{tab:ito}
  \begin{tabularx}{0.4\textwidth}{C|CCCC}
    \hline\hline
    \diagbox[width=2em]  & $d_x$  & $d_y$ & $d_w$  \\
    \hline
  GiTO$_9$      & $\sqrt{2}$   & $\sqrt{2}$ &  $2$      \\[1pt]
  GiTO$_{12}$   & $\sqrt{2}$    & $1$         & $\sqrt{2}$  \\[1pt]
  GiTO$_{12}'$   & $1$         & $\sqrt{2}$ &  $\sqrt{2}$  \\[1pt]
  GiTO$_{16}$    & $1$        & $1$      & $1$          \\
         \hline\hline
  \end{tabularx}
\end{table}

Now we apply the anyon condensation theory to the folded system with the $\Z_2^{\rm ex}$ symmetry included. That is, we set $\mathcal{P} = \mathcal{D}$ (see Sec.~\ref{sec:D}), and $\mathcal{U}$ is a topological order obtained by gauging the full $\Z_2^f \times \Z_2^{\rm ex}$ symmetry in the iTO. We aim to determine $\eta_\mathcal{M}$ from the data of $\mathcal{D}$ through anyon condensation theory. To find $\eta_\mathcal{M}$ through Eq.\eqref{eq:etam-def}, we need to determine $\theta_w$ and $\theta_x$. As Eq.~\eqref{eq:parityspin} already gives $\theta_w$, in this subsection we focus on how to determine $\theta_x$, the topological spin of a $\Z_2^{\rm ex}$ vortex.

\subsubsection{$\Z_2^f\times \Z_2^{\rm ex}$-gauged iTO}
\label{sec:z2z2ito}

Let us first discuss properties of $\mathcal{U}$, which we obtain by gauging $\Z_2^f\times \Z_2^{\rm ex}$ symmetric iTOs.  As discussed in the introduction,  2D $\Z_2^f\times \Z_2^{\rm ex}$ symmetric iTOs are classified by $\Z\times \Z_8$. According to Refs.~\cite{kitaev2006,GuLevin2014,WangPRB2017}, gauging $\Z^f_2\times\Z_2^{\rm ex}$ will give rise to $128$ gauged theories\footnote{The counting of distinct gauge theories depends on the criterion. We treat two gauge theories as topologically distinct, if there is no mapping between them such that properties of fusion and braiding and properties of gauge charge and flux are preserved. If only fusion and braiding properties are preserved, the counting of distinct GiTOs will be reduced. }. They are distinguished by the two quantities 
\begin{align}
\theta_{w} & =e^{i\pi \mu_1/8}, \nonumber\\
(\theta_{x})^2 & = e^{i\pi \mu_2/4},
\end{align}
where $w$ denotes a fermion parity vortex, $x$ denotes a $\Z_2^{\rm ex}$ vortex, and $\mu_1,\mu_2$ are integers discussed in Sec.~\ref{sec:eta}. Note that there may exist many fermion-parity vortices and $\Z_2^{\rm ex}$ vortices, but all of them give the same values of $\theta_w$ and $(\theta_x)^2$.

Depending on the number of anyons, we divide the 128 gauged theories into four categories
\begin{align}
{\rm GiTO}_{9}: & \ 1^\pm, f^\pm, x^\pm, y^\pm, w \nonumber\\
{\rm GiTO}_{12}: & \ 1^\pm, f^\pm, x^\pm, y^\pm,  y_f^\pm, w^\pm \nonumber \\ 
{\rm GiTO}_{12}': & \ 1^\pm, f^\pm, x^\pm, x_f^\pm, y^\pm, w^\pm \nonumber \\ 
{\rm GiTO}_{16}: & \ 1^\pm, f^\pm, x^\pm, x_f^\pm, y^\pm,  y_f^\pm, w^\pm, w_f^\pm \nonumber
\end{align}
where GiTO stands for ``gauged invertible topological order'' and the subscript is the number of anyons in the theory. Our notation is explained as follows.  The four charge excitations are the vacuum $1^+$,  the $\Z_2^{\rm ex}$ charge $1^-$, the pure fermion $f^+$, and the $\Z_2^{\rm ex}$-charged fermion $f^-$. We use ``$x$'' to denote a vortex carrying $\Z_2^{\rm ex}$ flux, use ``$w$'' to denote a fermion-parity vortex, and use ``$y$'' to denote a vortex carrying both $\Z_2^{\rm ex}$ and fermion-parity flux.  There exist several $x$-, $y$-, and $w$-vortices, differing by attaching charge excitations. Note that the sign ``$\pm$'' in our notation only has a relative meaning that vortices differ by a $\Z_2^{\rm ex}$ charge. All charges are Abelian anyons, while vortices may be non-Abelian. In a fixed GiTO, all $x$-vortices have the same quantum dimension, which we denote as $d_x$. Similarly, all $y$-vortices or $w$-vortices have the same quantum dimension, which we denote as $d_y$ and $d_w$ respectively.   Quantum dimensions for different GiTOs are listed in Table \ref{tab:ito}. By matching chiral central charge, one can show that GiTO$_9$ and GiTO$_{16}$ occur only for $c$ being even multiples of $1/4$, while  GiTO$_{12}$ and GiTO$_{12}'$ occur for $c$ being odd multiples of $1/4$, where $c$ is the central charge of $\mathcal{B}$.

We list a few fusion rules. Since fermion-parity vortices have been studied in the above section (Sec.~\ref{sec:ignorez2ex}), here we focus on the $x$- and $y$-vortices. First, we have
\begin{equation}
x^+\times 1^- = x^-, \quad y^{+}\times 1^{-} = y^-.
\end{equation}
Second, when $d_x = \sqrt{2}$, the two $x$-vortices $x^\pm$ satisfy
\begin{equation}
x^+ \times f^+ = x^-, \quad x^+ \times f^- = x^+.
\end{equation}
It is similar for $y$-vortices when $d_y = \sqrt{2}$. Third, when $d_x=1$, there are four $x$-vortices $x^\pm$ and $x_f^\pm$, and they satisfy
\begin{equation}
x^+ \times f^+ = x_{f}^-, \quad x^+ \times f^- = x_f^+.
\end{equation}
It is similar for Abelian $y$-vortices. 
 
The topological spins satisfy
\begin{equation}
\theta_{x^-} = - \theta_{x^+}, \quad \theta_{y^-} = - \theta_{y^+}.
\end{equation}
 In the presence of $x_f^\pm$ and/or $y_f^\pm$, the topological spins satisfy
\begin{align}
\theta_{x_f^+} & = \theta_{x^+} =  -\theta_{x_f^-},\nonumber\\
\theta_{y_f^+} & = \theta_{y^+} =  -\theta_{y_f^-} .
\end{align}
The first equality in each line is conventional: One can show that $\theta_{x^{+}}$ must be equal to either $\theta_{x_f^+}$ or $\theta_{x_f^-}$, and we use the convention that $x_f^+$ is the one that has the same topological spin as $x^+$. We also list some elements of the $S$ matrix of $\mathcal{U}$:
\begin{align}
S_{x^+, 1^+} &=S_{x^+, f^+}= -S_{x^+, 1^-} = -S_{x^+, f^-} = \frac{d_{x}}{D_\mathcal{U}}, \nonumber\\
S_{y^+, 1^+} &=-S_{y^+, f^+}= -S_{y^+, 1^-} = S_{y^+, f^-} = \frac{d_{y}}{D_\mathcal{U}},
\label{eq:smatrix-U}
\end{align}
where $D_\mathcal{U} = 4$ is the total quantum dimension (the same for all GiTOs).

\subsubsection{Anyon condensation}

Like in Sec.~\ref{sec:ignorez2ex}, our goal is to determine $\mathcal{U}$ out of the 128 possible GiTOs, provided that $\mathcal{P} = \mathcal{D}$ and  certain properties of the gapped domain are given.  Properties of anyons in $\mathcal{D}$ are given in Sec.~\ref{sec:D}. The fusion category $\mathcal{T}$ that lives on the domain wall contains the following anyons
\begin{equation}
\mathcal{T} = \{\underbrace{1^\pm, f^\pm, a^\pm, \dots}_{\text{anyons from } \mathcal{C}}, \underbrace{x_1^\pm, x_2^\pm, \dots}_{x\text{-vortices}}, \underbrace{y_1^\pm, y_2^\pm, \dots}_{y\text{-vortices}}, \underbrace{w_1, \dots}_{\text{others}}\}. \nonumber
\end{equation}
It is an extension of the fusion category $\mathcal{T}$ in Sec.~\ref{sec:ignorez2ex} by gauging $\Z_2^{\rm ex}$ symmetry. We do not aim to understand $\mathcal{T}$ completely. A partial understanding is enough for us to achieve our goal. 

Several properties of $\mathcal{T}$ are as follows. First, objects in $\mathcal{T}$ can be divided into those from $\mathcal{C}$ (with additional $\Z_2^{\rm ex}$ charge attached), $x$-vortices, $y$-vortices, and others (including $w$-vortices and $\Z_2^{\rm extra}$ vortices), similarly to those in $\mathcal{U}$. Second, every anyon $a\in \mathcal{C}$ is further decorated by a $\Z_2^{\rm ex}$ charge, denoted by ```$a^\pm$''. It is similar for the $x$- and $y$-vortices. However, $w$-vortices may not display this decoration as it may absorb the $\Z_2^{\rm ex}$ charge $1^-$. Third, the deconfined anyons form the category $\mathcal{U}$, discussed in Sec.~\ref{sec:z2z2ito}. We use the convention that $x_1^\pm \equiv x^\pm$ and $y^\pm_1 \equiv y^\pm$ are always deconfined $x$- and $y$-vortices. There may exist additional deconfined $x$- and $y$-vortices. Fourth, the topological spin of deconfined $w$-vortices is determined by \eqref{eq:parityspin} before gauging $\Z_2^{\rm ex}$, so we will not discuss it below. Its value does not change after we gauge $\Z_2^{\rm ex}$.

We first discuss the restriction maps. We claim that the restriction maps are given as follows:
\begin{subequations}
\label{eq:gr}
\begin{align}
r\{[a,b]\} & = \sum_c N_{a\rho_m(b)}^c(c^+ + c^-), \label{eq:gr1}\\
r\{(a,a)^+\} & = \sum_c N_{a\rho_m(a)}^{c} c^{\mu[a,\rho_m(a);c]}, \label{eq:gr2}\\
r\{(a,a)^-\} & = \sum_c N_{a\rho_m(a)}^{c} c^{-\mu[a,\rho_m(a);c]}, \label{eq:gr3}\\
r\{[v_1,v_2]\} & = w\text{-vortices only}, \label{eq:gr4} \\
r\{(v,v)^\pm\} & = w\text{-vortices only}, \label{eq:gr5} \\
r\{[a,v]\}& = \text{confined anyons only}, \label{eq:gr6} \\
r\{X_a^\pm\} & = \text{confined anyons only}, \label{eq:gr7} \\
r\{X_v^\pm\}& = \text{$x$- and $y$-vortices only}, \label{eq:gr8}
\end{align}
\end{subequations}
where we use $r\{\cdot\}$ to denote restriction maps for notational clarity. They are some kind of extensions of the restriction maps in \eqref{eq:r}. The restriction maps \eqref{eq:gr1}-\eqref{eq:gr3} can be inferred from \eqref{eq:r1} and the mirror symmetry properties of the anyons, as discussed in Sec.~\ref{sec:folding-gauged}. The sign $\mu[a,\rho_m(a);c]$ is not known generally. However, if $c=1$ or $f$, we must have $\mu[a,\rho_m(a);c] = \mu_a$, the mirror eigenvalue defined in Eq.~\eqref{eq:mirroreig}. Special cases of Eqs.~\eqref{eq:gr1}-\eqref{eq:gr3} are
\begin{align}
r\{(1,1)^\pm\} & = 1^\pm, \nonumber\\
r\{(f,f)^\pm\} & = 1^\mp, \nonumber\\
r\{[1,f]\} & = f^+ + f^-. 
\end{align}
The correlation of ``$\pm$'' signs between \eqref{eq:gr2} and \eqref{eq:gr3} follows from the relation that $r\{(a,a)^+\}\times r\{(1,1)^-\} = r\{(a,a)^-\}$. The restriction maps \eqref{eq:gr4}-\eqref{eq:gr6} follow from \eqref{eq:r2}-\eqref{eq:r4}. Since we are only interested in the deconfined $x$- and $y$-vortices in this subsection, these restriction maps are unimportant.  To understand the map \eqref{eq:gr7}, we consider the mutual statistics between $(f,f)^-$ and $X_\alpha^\pm$, which is
\begin{equation}
M_{(f,f)^-, X_\alpha^\pm} = - M_{f,\alpha} = \left\{
\begin{array}{ll}
-1, & \text{if $\alpha\in \mathcal{C}$} \nonumber\\[3pt]
1, & \text{if $\alpha\in \bar{\mathcal{C}}$}
\end{array}
\right.
\end{equation}
which follows from the $S$ matrix \eqref{eq:smatrix-of-D} and the fact that $(f,f)^-$ is Abelian. Since $(f,f)^-$ becomes the vacuum $1^+$ in $\mathcal{U}$,  $X_a^\pm$ restricts to confined anyons only. In fact, $(f,f)^-$ is the condensed charge of the $\Z_2^{\rm extra}$ symmetry. Then, $X_v^\pm$ restricts to $x$- and $y$-vortices only, giving rise to \eqref{eq:gr8}.

The restriction map \eqref{eq:gr8} is an important one. We expand it in more detail:
\begin{align}
r\{X_v^+\} & = n_v x^+ + p_v y^+ + ( n_{v}'x_f^+ + p_{v}'y_f^+ ) + \dots \label{eq:gr-vortex}
\end{align}
where $n_v, p_v,  n_{v}',  p_{v}'$ are integers (to be determined), and ``$\dots$'' are confined anyons. Only vortices with ``$+$'' sign appear on the right-hand side, because all deconfined anyons must have the same topological spins and vortices with different signs cannot have equal topological spin due to our convention.  The vortices $x_f^+$ and/or $y_f^+$ may or may not appear in \eqref{eq:gr-vortex}, depending on the details of the condensation, and thereby we put a parenthesis around them. The restriction map $r\{X_v^-\}$ can be induced from  $r\{X_v^+\}$, using the fact $r\{X_v^+\}\times r\{(1,1)^-\} = r\{X_v^-\}$. Therefore, we have
\begin{align}
r\{X_v^-\} & = n_v x^- + p_v y^- + ( n_{v}'x_f^- + p_{v}'y_f^- ) + \dots 
\end{align}
In addition, from the commutativity \eqref{eq:rr=r} with $\alpha =X_v^+$ and  $\beta=[1, f]$, we obtain
\begin{align}
 r\{X_{vf}^+\} + r\{X_{vf}^-\}= r\{X_v^+\}\times f^+ + r\{X_v^+\}\times f^-,
\label{eq:fusef}
\end{align}
where the subscript $vf$ should be identified to $v$ if it is of Majorana type. Expanding the restriction maps in \eqref{eq:fusef} for various GiTOs, we obtain the following relations
\begin{align}
{\rm GiTO}_{9}: & \ n_{vf}= n_v,\ p_{vf} = p_v \nonumber\\
{\rm GiTO}_{12}: & \ n_{vf}=n_v, \ p_{v}' = p_{vf}, \ p_{vf}'=p_v \nonumber \\ 
{\rm GiTO}_{12}': & \ n_v'=n_{vf}, \ n_{vf}'=n_v, \ p_v= p_{vf}\nonumber \\ 
{\rm GiTO}_{16}: & \ n_v'=n_{vf}, \ n_{vf}'=n_v,  \ p_{v}' = p_{vf}, \ p_{vf}'=p_v
\label{eq:hatnp-prop}
\end{align}
Accordingly, $\{n_v'\}$ and $\{p_v'\}$, if needed, are completely determined by $\{n_v\}$ and $\{p_v\}$. Hence, only $\{n_v,p_v\}$ are independent in the restriction maps. In addition, we may have the constraint $n_v=n_{vf}$ and/or $p_v=p_{vf}$, depending on the scenario.

From the restriction maps, it is straightforward to read out the lifting maps for the deconfined anyons in $\mathcal{U}$. We have
\begin{subequations}
\label{eq:gl}
\begin{align}
l(1^\pm) & = \!\sum_{a=\bar\rho_m(a)} \!(a, a)^{\pm\mu_a} + \!\!\sideset{}{'}\sum_{a\neq\bar \rho_m(a)}\! \left[a, \bar{\rho}_m(a)\right] , \label{eq:gl1}\\
l(f^\pm) & = \!\!\sum_{af = \bar\rho_m(a)}\!\!\! (a, a)^{\pm\mu_a} + \!\!\!\sideset{}{'}\sum_{af \neq \bar\rho_m(a)} \!\!\! \left[af, \bar{\rho}_m(a)\right], \label{eq:gl2}\\
l(x^\pm) & = \sum_{v\in\bar{\mathcal{C}}} n_v X_{v}^\pm ,\label{eq:gl3}\\
l(y^\pm) & = \sum_{v\in\bar{\mathcal{C}}} p_v X_{v}^\pm , \label{eq:gl4}\\
l(x_f^\pm) & = \sum_{v\in\bar{\mathcal{C}}} n_{v}' X_{v}^\pm,  \label{eq:gl5}\\
l(y_f^\pm) & = \sum_{v\in\bar{\mathcal{C}}} p_v' X_{v}^\pm,  \label{eq:gl6}
\end{align}
\end{subequations}
where ``$\sum'$'' means that it sums only one anyon in $\{a,\bar\rho_m(a)\}$ or ``$\{af,\bar \rho_m(a)\}$'', when the two are not the same. The ``$\pm$'' signs in all the lifting maps are correlated on two sides of the equations. The lifting maps \eqref{eq:gl5} and \eqref{eq:gl6} may or may not be needed, depending on details of the condensation. We will say $X_{v}^+ \in l(x^+)$  if $n_v \neq 0$, and similarly for other lifting maps.

%Generally speaking, the restriction and lifting maps may not fully characterize anyon condensation. However, for our purpose of determining the mirror anomaly $\eta_{\mathcal{M}}$, they already provide a lot of information. 

\begin{table*}
  \caption{Some entries of the matrix $n_{\alpha,t}$ for $\alpha\in\mathcal{D}$ and $t\in\mathcal{U}$.}
  \label{tab:nat}
  \begin{tabularx}{0.9\textwidth}{C|CCCCCC CCCC CC}
    \hline\hline
    \diagbox[width=4em]{$\alpha$}{$t$}& $1^+$ &  $1^-$ & $f^+$ & $f^-$ & $x^+$  & $y^+$ & $x_f^+$ & $y_f^+$\\
    \hline
    $(a,a)^{\mu_a}$  & $\delta_{a,\bar\rho_m(a)}$ & 0 & $\delta_{af, \bar\rho_m(a)}$ & 0 & 0 & 0 & 0 & 0\\[1pt]
    $(a,a)^{-\mu_a}$ & 0 & $\delta_{a,\bar\rho_m(a)}$ & 0 & $\delta_{af, \bar\rho_m(a)}$ & 0 & 0 & 0 & 0\\[1pt]
    $[a,b]$      & $\delta_{b,\bar\rho_m(a)}$ & $\delta_{b,\bar\rho_m(a)}$  & $\delta_{bf,\bar\rho_m(a)}$ & $\delta_{bf,\bar\rho_m(a)}$ & 0 & 0  & 0 & 0\\[3pt]
    $X_v^+$           & 0 & 0 &  0 & 0 &$n_v$ & $p_v$ & $n_v'$ & $p_v'$ \\[1pt]
%    $X_v^-$           & 0 & 0 & 0 & 0 & 0 & $n_v$ \\
    \hline\hline
  \end{tabularx}
\end{table*}

\subsubsection{Relation between $\{\mu_a\}$ and $\{n_v, p_v\}$ }

The original mirror SETs are characterized by $\rho_m$ and $\{\mu_a\}$. Above, we have defined the lifting/restriction maps, and introduced the integers $\{n_v,p_v\}$, as well as $\{n_v', p_v'\}$. The latter integers $\{n_v', p_v'\}$ are determined by $\{n_v,p_v\}$ through \eqref{eq:hatnp-prop}. These integers are so far unknown. Below we show that the two sets of data, $\{\mu_a\}$ and $\{n_v,p_v\}$, are closely related. 

We describe the relation by answering the following two questions:
\begin{enumerate}
\item How do we determine if  we need $\{n_v'\}$ and/or $\{p_v'\}$? 
\item How do we determine the values of $\{n_v,p_v\}$, provided that $\{\mu_a\}$ are given? 
\end{enumerate}
To answer the first question, it is enough to compute $d_x$ and $d_y$, as the two quantum dimensions determine the GiTO type of $\mathcal{U}$ as shown in \ref{tab:ito}. The type of $\mathcal{U}$ determines if  we need $\{n_v'\}$ and/or $\{p_v'\}$. To the second question, we find that, instead of $\{n_v, p_v\}$, it is easier to determine $\{\hat{n}_v, \hat p_v\}$, which are defined as
\begin{equation}
\hat{n}_v =\left\{
\begin{array}{ll}
n_v, &\quad \text{if } d_x= \sqrt{2}\\
n_v+n_{vf}, & \quad \text{if } d_x =1
\end{array}
\right. 
\label{eq:hatn-def}
\end{equation}
and
\begin{equation}
\hat{p}_v =\left\{
\begin{array}{ll}
p_v, &\quad \text{if } d_y= \sqrt{2}\\
p_v+p_{vf}, & \quad \text{if } d_y =1
\end{array}
\right. 
\label{eq:hatp-def}
\end{equation}
Note that the two definitions are meaningful only after we have obtained $d_x$ and $d_y$. Again, it is understood that $vf\equiv v$ if $v$ is a Majorana-type vortex. Unfortunately, we do not know how to determine $\{n_v, p_v\}$ from $\{\hat n_v, \hat p_v\}$ in general; we will comment on this in Sec.~\ref{sec:add_prop} through an example. Nevertheless, the knowledge of $\{\hat{n}_v, \hat p_v\}$ will be enough for us to derive the anomaly indicator, as we will show in Sec.~\ref{sec:indicator}.

Now we claim that (partial) answers to the above two questions are implied by the following relations,
\begin{align}
\left\{\begin{matrix}
\displaystyle
\sum_{[a]}\Lambda_{a,v}\mu_a = \frac{d_x\hat n_v}{\sigma_v},  \\[5pt]
\displaystyle
\sum_{[a]}\Lambda_{a,v}\mu_a \xi_a= \frac{d_y\hat p_v}{\sigma_v},
\end{matrix}
\right. 
\label{eq:Lambda-relation1}
\end{align}
and its inverse
\begin{align}
\left\{\begin{matrix}
\displaystyle
\sum_{[v]} \Lambda_{a,v} \left(\frac{d_x\hat n_v}{\sigma_v}\right) &  = \mu_a, \\
\displaystyle
\sum_{[v]} \Lambda_{a,v} \left(\frac{d_y\hat p_v}{\sigma_v}\right)&  = \mu_a\xi_a,\end{matrix}
\right.
\label{eq:Lambda-relation2}
\end{align}
where summations are over all pairs $[\alpha] =\{\alpha,\alpha f\}$ (with $\alpha = a$ or $v$), or equivalently, only one anyon in each pair is summed over. One can check such summations are well-defined (i.e., independent of which anyon to use  in each pair).   The left- and right-hand sides are inverse transformations of each other, since $\Lambda$ can be viewed as a unitary matrix when its indices only go through the pairs $\{[\alpha]\}$ (see Appendix \ref{sec:app_B_prop} for properties of the $\Lambda$ matrix).

Before deriving \eqref{eq:Lambda-relation1} and \eqref{eq:Lambda-relation2}, let us see why they imply answers to the above two questions. First, they determine $d_x$ and $d_y$. We assume that $\rho_m$ and $\{\xi_a,\mu_a\}$ of mirror SETs, and $\Lambda$ and $\{\sigma_v\}$ of $\mathcal{B}$ are given. Plugging into \eqref{eq:Lambda-relation1}, we claim that the left-hand side must be integer or multiples of $\sqrt{2}$, in order to be equal to the right-hand side; if not, the original SET data $\rho_m$ and $\{\mu_a\}$ are problematic. Since $\{\hat n_v, \hat p_v\}$ are integers, whether $d_x$ and $d_y$ are 1 or $\sqrt{2}$ can be immediately computed. Once $d_x$ and $d_y$ are known, the quantities $\{\hat{n}_v\}$ and $\{\hat p_v\}$ are well-defined through \eqref{eq:hatn-def} and \eqref{eq:hatp-def}. Then, we can compute $\{\hat{n},\hat{p}\}$ from $\{\mu_a\}$ by \eqref{eq:Lambda-relation1}.

The derivation of \eqref{eq:Lambda-relation1} and \eqref{eq:Lambda-relation2} can be done in a case-by-case way for GiTO$_9$, GiTO$_{12}$, GiTO$_{12}'$ or GiTO$_{16}$ separately. The procedure is very similar in each case. Here we show the derivation in the case that $\mathcal{U}$ is GiTO$_9$, i.e., $d_x=d_y=\sqrt{2}$. Taking $\alpha= X_v^+ \in l(x^+)$ and $t = 1^+$ in Eq.~\eqref{eq:sn=ns},  inserting the restriction and lifting coefficients in \eqref{eq:gr} and \eqref{eq:gl},  and using the $S$ matrix elements in Eqs.~\eqref{eq:smatrix-of-D} and \eqref{eq:smatrix-U}, we obtain
\begin{equation}
\sum_{a = \bar\rho_m(a)}S_{a,v}\mu_a = \frac{d_x}{2} n_v +\frac{d_y}{2}p_v .
\label{eq:munp1}
\end{equation}
Similarly, if we take $\alpha= X_v^+ \in l(x^+)$ and $t = f^+$ in Eq.~\eqref{eq:sn=ns}, we obtain
\begin{equation}
\sum_{af = \bar\rho_m(a)}S_{a,v}\mu_a = \frac{d_x}{2} n_v -\frac{d_y}{2}p_v .
\label{eq:munp2}
\end{equation}
Then, combining \eqref{eq:munp1} and \eqref{eq:munp2} and using the definitions \eqref{eq:mirroreig2} and \eqref{eq:xi}, we  immediately obtain \eqref{eq:Lambda-relation1}. The same results can be obtained if we take $\alpha = X_v^-$ and $t=1^-, f^-$  in \eqref{eq:sn=ns}. Instead, if $\alpha = (a,a)^+$ and $t=x^+, y^+$, with some straightforward computations, we immediately obtain \eqref{eq:Lambda-relation2}. 

\subsubsection{Additional properties}
\label{sec:add_prop}

The relations \eqref{eq:Lambda-relation1} and \eqref{eq:Lambda-relation2} are very useful. They are consequences of the constraint \eqref{eq:sn=ns}, and can be used to determine $\{\hat{n}_v, \hat{p}_v\}$ from $\{\xi_a, \mu_a\}$, and vice versa. However, they are not all the constraints. Equation \eqref{eq:rr=r} is very restrictive and we have not used its full power. For example, by considering $\alpha = X_v^\pm$ and  $\beta = X_{v'}^\pm$ or $[a,b]$ in \eqref{eq:rr=r}, one can further constrain $\{n_v,p_v\}$ beyond the simple relations \eqref{eq:hatnp-prop}. Combining with \eqref{eq:Lambda-relation1}, one may be able to determine $\{n_v, p_v\}$, instead of just $\{\hat{n}_v, \hat{p}_v\}$, from $\{\xi_a, \mu_a\}$. In other cases, we may not know what are the valid choices of $\{\mu_a\}$, given a permutation $\rho_m$. Then, if we are able to find constraints on $\{n_v, p_v\}$, then they can be used to constrain $\{\mu_a\}$
through \eqref{eq:Lambda-relation2}. 

However, we will not explore the full power of \eqref{eq:rr=r} generally. On the one hand, it requires more knowledge of the $\Z_2^f$-gauged theory $\mathcal{B}$, e.g., fusion rules of fermion-parity vortices,  which are not generally known. On the other hand, \eqref{eq:Lambda-relation1} and \eqref{eq:Lambda-relation2} are enough for us to derive the expression \eqref{eq:etam} for the anomaly indicator $\eta_{\mathcal{M}}$. In Sec.~\ref{sec:sf2}, we will discuss one example to showcase that additional constraints on $\{n_v, p_v\}$ can be obtained from \eqref{eq:rr=r}.

Here, we show one general property obtained from \eqref{eq:rr=r}. We claim that
\begin{align}
\sum_{v\in\bar{\mathcal{C}}} \hat{n}_v d_v & = D_\mathcal{B}/d_x, \nonumber\\
\sum_{v\in\bar{\mathcal{C}}} \hat{p}_v d_v & = D_\mathcal{B}/d_y.
\label{eq:nd-dimension}
\end{align}
To derive this result, we make use of a corollary of \eqref{eq:rr=r}
\begin{equation}
q d_t = \sum_{\alpha\in\mathcal{P}} n_{\alpha, t} d_\alpha,
\label{eq:qdt}
\end{equation}
where $t\in\mathcal{U}$ and $q = \sum_{\alpha\in\mathcal{P}} n_{\alpha, 1} d_\alpha$ (e.g., see Ref.~\onlinecite{NeupertPRB2016} for a derivation). Taking  $\mathcal{P}=\mathcal{D}$ and using the lifting map \eqref{eq:gl1},  we find
\begin{equation}
q = D_\mathcal{C}^2 = D_\mathcal{B}^2/2 .
\label{eq:q1}
\end{equation}
Applying \eqref{eq:qdt} with the lifting maps \eqref{eq:gl3} and \eqref{eq:gl4}, we obtain
\begin{align}
q d_x = D_\mathcal{B} \sum_{v}n_v d_v, \quad q d_y = D_\mathcal{B} \sum_{v}p_v d_v
\label{eq:q2}
\end{align}
where we used $d_{X_v^\pm} = d_vD_\mathcal{B}$. Combining \eqref{eq:q1} and \eqref{eq:q2} with the definitions \eqref{eq:hatn-def} and \eqref{eq:hatp-def}, we immediately obtain \eqref{eq:nd-dimension}.

%We would like to distinguish the pure $\Z_2^M$ defects in $l(x_1^\pm)$ and those carrying also fermion parity flux, i.e., those in $l(x_v^\pm)$. To do that, we need to condense $(f,f)^-$ so that $[1, f] \rightarrow f^+ + f^-$, making the fermion  Abelian. First of all, we have
%\begin{equation}
%X_\alpha^+ \times (f,f)^- = X_\alpha^- 
%\end{equation}
%This means $X_\alpha^\pm$ does not split. and 
%\begin{equation}
%M_{X_\alpha^\pm, (f,f)^-} = -S_{\alpha,b}
%\end{equation}
%For $X_\alpha^\pm$ to be deconfined, we must have $\alpha$ to be an fermion parity flux. That is, only $X_\alpha^\pm$ with $\alpha$ being a parity flux can be $\Z_2^M$ fluxes. 

%We consider the fusion rule for arbitrary $v$
%\begin{equation}
%X_v^+\times X_v^+ = 
%\end{equation}
%To know this, we will need to know the fusion rule of fermion-parity vortices. Unfortunately, it is not generally known yet about the fusion rules of $\mathcal{B}$ (not simple algorithm known yet).

\section{Anomaly indicator}
\label{sec:indicator}

In this section, we derive the formula \eqref{eq:etam} for the anomaly indicator $\eta_\mathcal{M}$, i.e., express $\eta_\mathcal{M}$ in terms of topological and symmetry properties of $\mathcal{C}$.  According to the definition \eqref{eq:etam-def},  $\eta_\mathcal{M}$ is determined by two topological spins, $\theta_w$ and $(\theta_{x^+})^2$. The spin $\theta_w$ is expressed in terms of the chiral central charge $c$ of $\mathcal{C}$ in  \eqref{eq:parityspin}.  The spin $\theta_{x^+}$ is equal to $\theta_{X_v^+}$ for any $X_v^+\in l(x^+)$, i.e., for any $v$ with $n_v\neq 0$. With the expression \eqref{eq:vortexspin},  we immediately have
\begin{align}
\eta_{\mathcal{M}} = \theta_{w}(\theta_{X_v^+}^*)^2  = e^{i2\pi c/8}\theta_v^*.
\label{eq:etam-alt}
\end{align}
We emphasize that for all $v$'s with $n_v\neq 0$, $X_v^+$ must have the same topological spin. We note that \eqref{eq:etam-alt}  expresses $\eta_\mathcal{M}$ in terms of quantities in $\mathcal{B}$.

Now we claim that the formulas \eqref{eq:etam} and \eqref{eq:etam-alt} are equivalent. To prove the claim, we multiply Eq.~\eqref{eq:etam} on both sides by $e^{-i2\pi c/8}$ and do the following steps:
\begin{align}
\eta_{\mathcal{M}} e^{- i2\pi c/8} & = \frac{1}{D_\mathcal{B}^2}\sum_{a\in \mathcal{C}} d_a\theta_a \mu_a \sum_{\alpha\in\mathcal{B}}d_\alpha^2\theta_\alpha^*\nonumber\\
&  = \frac{1}{D_\mathcal{B}^2} \sum_{a, \alpha} \theta_a \mu_a \theta_\alpha^* d_\alpha \sum_{\beta\in\mathcal{B}} N_{a\alpha}^{\beta} d_{\beta}\nonumber\\
& = \frac{1}{D_\mathcal{B}} \sum_{a,\beta} \mu_a d_\beta \theta_{\beta}^*\left(\frac{1}{D_\mathcal{B}}  \sum_\alpha\frac{\theta_a\theta_{\beta}}{\theta_\alpha} N_{a\bar{\beta}}^{\bar\alpha} d_\alpha\right) \nonumber\\
& = \frac{1}{D_\mathcal{B}} \sum_{a,\beta} \mu_a d_\beta \theta_{\beta}^* S_{a,\beta}^* \nonumber\\
& =  \frac{1}{D_\mathcal{B}} \sum_{v\in\bar{\mathcal{C}}} d_v \theta_v^* \frac{\sigma_v}{2}\sum_{a} \mu_a \Lambda_{a,v}^* \nonumber\\
& =  \frac{1}{D_\mathcal{B}} \sum_{v} d_v \theta_v^* \hat{n}_v d_x  \nonumber\\
& =  \frac{1}{D_\mathcal{B}}  \theta_v^*d_x \sum_{v} d_v \hat{n}_v  \nonumber\\
& = \theta_v^*,
\label{eq:eta_dev}
\end{align}
where $v$ in the last equation can be any fermion parity vortex with  $n_v\neq 0$. In the first line, we have inserted the formula \eqref{eq:centralcharge} and used the relation $D_\mathcal{B} = \sqrt{2}D_\mathcal{C}$. In the second line, the relation $d_a d_\alpha = \sum_{\beta} N_{a\alpha}^\beta d_\beta$ is inserted. In the third line, we have used $N_{a\alpha}^\beta = N_{a\bar\beta}^{\bar \alpha}$. The term inside the parenthesis is the definition of $S_{a,\beta}$, which leads to the forth line. In the fifth line, we have reduced the summation over $\beta$ to summation over vortices $v$ only, as summing over $a\in\mathcal{C}$ is 0. We have also inserted the definition of $\Lambda$ matrix. In the sixth line, we have used Eq.~\eqref{eq:Lambda-relation1}. In the seventh line, we have used the fact that $\theta_v$ are all equal as long as $n_v\neq 0$. Finally, we use the relation \eqref{eq:nd-dimension}. Equation \eqref{eq:eta_dev} is equivalent to Eq.~\eqref{eq:etam-alt}. Hence, we prove our claim. 

This equivalence shows that \eqref{eq:etam} is indeed a formula for the anomaly indicator. The above derivation is adapted from a very similar derivation of anomaly indicator in the bosonic case\cite{folding}.

One may observe that the $x$- and $y$-vortices, as well as $\{n_v\}$ and $\{p_v\}$, appear in parallel in the anyon condensation formulation. Indeed, we may define an alternative mirror symmetry, $\mathcal{M}'=\mathcal{M}P_f$, which corresponds to the $y$-vortices after folding and gauging. The anomaly indicator \eqref{eq:etam} should work equally well for $\mathcal{M}'$. It is not hard to see that the new mirror eigenvalue $\mu_a' =\mu_a\xi_a$. Then, we have
\begin{equation}
\eta_{\mathcal{M}'}= \frac{1}{\sqrt{2}D_\mathcal{C}}\sum_{a\in\mathcal{C}} d_a\theta_a\mu_a\xi_a.
\label{eq:etamprime}
\end{equation}
This expression can also be derived  by replacing $x$-vortices with $y$-vortices in \eqref{eq:etam-alt} and following similar steps as in \eqref{eq:eta_dev}. The two indicators are not independent. One can show that $\eta_{\mathcal{M}'}=\eta_{\mathcal{M}}^*$. This relation can be proven either by directly comparing the expressions \eqref{eq:etam} and \eqref{eq:etamprime}, or by  using the definition \eqref{eq:etam-def} and checking the topological spins of the $x$-, $y$-, $w$-vortices in the GiTOs.

%The problem is that here $\beta$ can be any anyon. What if $%\beta\in A$, we have
%\begin{align}
%\sum_a \mu(a) S_{a,b} & = \sum_{a} \mu(a) S_{a,b} + \mu(af) %S_{af,b} \nonumber\\
%& = \sum_{a} \mu(a) S_{a,b} - \mu(a) S_{a,b} =0
%\end{align}
%So, the summation over $\beta$ is actually only over the %parity fluxes. Then, we have
%\begin{equation}
%\eta = e^{i2\pi c_-/8} \theta_v^* = e^{i2\pi c_-/4}\Theta_1^* %= \Theta_0 \Theta_1^*
%\end{equation}
%This is exactly the needed anomaly indicator.

\section{Examples}
\label{sec:examples}

In this section, we give a few explicit examples, with an emphasis on the application of the general procedure discussed in Sec.~\ref{sec:anyoncon}. Most examples that we discuss are originally proposed in time-reversal symmetric systems. However, they can be easily adapted into mirror-symmetric topological orders. 

\subsection{Semion-fermion topological order}
\label{sec:sf}

\begin{table}
\caption{Data for $\rm SF_{\kappa}$ with $\kappa = \pm 1$. Lifting coefficients for both Abelian and non-Abelian $\mathcal{B}$ are listed.}
\label{tab:sf}
\begin{tabular}{c|cccc}
\multicolumn{5}{c}{SET data}\\
\hline 
 & $\ 1\ $ & $\ f\ $ & $\ s\ $ & $\ sf\ $ \\
\hline 
$\ \mu_a\ $ & $1$ & $-1$ & $\kappa$ & $-\kappa$ \\
$\xi_a$ & $1$ & $1$ & $-1$ & $-1$\\
\hline 

\end{tabular}
\hspace{15pt}
\begin{tabular}{c|cc}
\multicolumn{3}{c}{non-Abelian $\mathcal{B}$; $ d_x=d_y=1$}\\
\hline
 & $\ \tau\ $  & $\ \tau s\ $  \\
\hline 
$\sigma_v$ & $\sqrt{2}$ & $\sqrt{2}$ \\
$  n_v$ & $\ (1+\kappa)/2\ $ & $\ (1-\kappa)/2 \ $   \\
$  p_v$ & $\ (1-\kappa)/2\ $ & $\ (1+\kappa)/2 \ $ \\
\hline
\end{tabular}
\bigskip
\vspace{0pt}
\begin{tabular}{c|cccc}
\multicolumn{5}{c}{$\ $ Abelian $\mathcal{B}$; $ d_x=d_y=\sqrt{2}$}\\
\hline
 & $\ \tau\ $ & $\ \tau f\ $ & $\ \tau s\ $ & $\ \tau sf\ $ \\
\hline 
$\sigma_v$ & $1$ & $1$ & $1$ & $1$\\
$  n_v$ & $\ (1+\kappa)/2\ $ & $\ (1+\kappa)/2\ $ & $\ (1-\kappa)/2\ $  & $\ (1-\kappa)/2\ $  \\
$  p_v$ & $\ (1-\kappa)/2\ $ & $\ (1-\kappa)/2\ $ & $\ (1+\kappa)/2\ $  & $\ (1+\kappa)/2\ $ \\
\hline 
\end{tabular}
\end{table}

The semion-fermion (SF) topological order is the simplest nontrivial fermionic topological order. It contains four anyons, 
\begin{equation}
\mathcal{C}_{\rm SF} = \{1, f, s, sf\}, \nonumber
\end{equation}
where $f$ is the transparent fermion, and $s$ is a semion with $\theta_s = i$. It is an Abelian topological order. One of the fusion rules is $s\times s = 1$ and others can be easily deduced. The only possible permutation by the mirror symmetry is as follows
\begin{equation}
\rho_m(1) = 1, \ \rho_m(f) = f, \ \rho_m(s) = sf, \ \rho_m(sf) = s.
\label{eq:sf-permutation}
\end{equation}
Under this permutation, there are two possible mirror symmetry fractionalizations, characterized by $\mu_s = \kappa$, with $\kappa = \pm 1$. Data for the values of $\{\mu_a\}$ and $\{\xi_a\}$ are summarized in Table \ref{tab:sf}. These two mirror-enriched SF topological orders are referred to as ``$\rm SF_+$'' and ``$\rm SF_-$'', respectively. 

There are sixteen possible $\Z_2^f$-gauged theories. Among them, let us first consider the Abelian ones; the non-Abelian ones will be discussed at the end. The Abelian $\Z_2^f$-gauged theories all contain 8 anyons: 
\begin{align}
\mathcal{B} &  = \{1, f, s, sf, \tau, \tau f, \tau s, \tau sf\} \nonumber\\
& =\{1, f, \tau, \tau f\} \boxtimes \{1, s\},
\nonumber
\end{align}
where $\tau$ denotes a fermion-parity vortex. All anyons are Abelian. The labelling scheme follows their fusion rules. The total quantum dimension $D_\mathcal{B} = 2\sqrt{2}$.  The eight possibilities are distinguished by the topological spin $\theta_\tau = e^{i2\pi z /8}$, with $z=0, 1, \dots, 7$. There are four fermion-parity vortices $\tau, \tau f, \tau s $ and $ \tau s f$, all of which are of non-Majorana type.  From Eq.~\eqref{eq:centralcharge}, we can compute the chiral central charge  of $\mathcal{B}$, which is given by
\begin{equation}
c = z + 1 \quad \modulo{8}.
\end{equation}
Since we only care about $c$ modulo 8, this equation is enough.  Mutual statistics can be determined by the formula $M_{\alpha,\beta}=\theta_{\alpha}\theta_\beta\theta_{\alpha\times\beta}$, where the fusion product $\alpha\times \beta$ contains a unique anyon since all anyons are Abelian. With $\mathcal{B}$,  we can obtain the double-layer topological order $\mathcal{B}\boxtimes \mathcal{B}$, and the topological order $\mathcal{D}$ after further gauging $\Z_2^{\rm ex}$, for which we refer the readers to the general discussions in Sec.~\ref{sec:double}.

Now we make use of the relations in \eqref{eq:Lambda-relation1} and \eqref{eq:Lambda-relation2} to determine the topological order $\mathcal{U}$. The matrix $\Lambda_{a,v}$ can be read out from the mutual statistics between anyons. Taking $a=\{1,s\}$ and $v=\{\tau,\tau s\}$, we write $\Lambda_{a,v}$ in a matrix form
\begin{equation}
\Lambda = \frac{1}{\sqrt{2}}\left(
\begin{matrix}
1 & 1 \\
1 & -1
\end{matrix}
\right).
\label{eq:SF-Lambda}
\end{equation}
Applying \eqref{eq:Lambda-relation1} with $\mu_a$, $\xi_a$, and $\sigma_v$ in Table \ref{tab:sf}, we obtain 
\begin{align}
(1+\kappa, 1-\kappa)/\sqrt{2} & = (\hat{n}_\tau, \hat{n}_{\tau s}) d_x, \nonumber\\
(1-\kappa, 1+\kappa)/\sqrt{2} & = (\hat{p}_\tau, \hat{p}_{\tau s}) d_y.
\end{align}
We must have $d_x = d_y = \sqrt{2}$ to have integer solutions for $\hat n$ and $\hat{p}$. Hence, $\mathcal{U}$ must be a GiTO$_9$ topological order. 

Specializing to $\rm SF_+$ with $\kappa = + 1$,  we have $\hat{n}_\tau = \hat{p}_{\tau s} = 1$ and $\hat{n}_{\tau s} = \hat{p}_{\tau } = 0$. With the definitions \eqref{eq:hatn-def} and \eqref{eq:hatp-def} of $\{\hat{n}_v\}$ and $\{\hat p_v\}$, and properties in \eqref{eq:hatnp-prop}, we have $n_\tau=n_{\tau f}=p_{\tau s} = p_{\tau s f} =1$ and $n_{\tau s} = n_{\tau s f} = p_{\tau} = p_{\tau f}=0$. Then, the lifting maps are given by
\begin{align}
l(1^+) & = (1,1)^+ + (f,f)^+ + [s, sf], \nonumber\\
l(f^+) & = [1, f] + (s,s)^+  + (sf,sf)^-, \nonumber\\
l(x^+) & = X_\tau^+ + X_{\tau f}^+, \nonumber\\
l(y^+) & = X_{\tau s}^+ + X_{\tau s f}^+. 
\end{align}
The lifting maps for $1^-, f^-, x^-$ and  $y^-$ are similar. We have the topological spin 
\begin{equation}
\theta_{x^+} =\theta_{X^+_\tau} = e^{i2\pi c/16  }\sqrt{\theta_\tau} = e^{i2\pi(2z+1)/16}.
\end{equation}
The anomaly indicator is given by $\eta_\mathcal{M} = e^{i2\pi c/4} (\theta_{x^+}^* )^2= e^{i\pi/4}$, independent of $z$. This shows that $\rm SF_+$ has an index $\nu=2$, i.e., it lives on the surface of  a 3D TCSC with index $\nu=2$. Of course, one may directly apply the formula \eqref{eq:etam} to obtain the same result. Similar calculations can be done for $\rm SF_-$, which is also anomalous and corresponds to an index $\nu = 14$.

Above, we have considered only the cases that the $\Z_2^f$-gauged theory $\mathcal{B}$ is Abelian. One may also consider the remaining 8 possible non-Abelian $\mathcal{B}$'s. They are given by
\begin{align}
\mathcal{B} &  = \{1, f, s, sf, \tau, \tau s\} \nonumber\\
& =\{1, f, \tau\} \boxtimes \{1, s\},
\nonumber
\end{align}
where $\tau$ is non-Abelian with $d_\tau = \sqrt{2}$. The vortex $\tau$ is Majorana-type. The topological spin $\theta_{\tau} = e^{i2\pi z/8}$, with $z = 1/2, 3/2, \dots, 15/2$. The matrix $\Lambda$ is again given by \eqref{eq:SF-Lambda}. Similar calculations can be done as above. One finds that $d_x=d_y=1$, i.e., $\mathcal{U}$ is a GiTO$_{16}$. Other quantities can be straightforwardly obtained, which are listed in Table \ref{tab:sf}. As expected, we again find that $\eta_\mathcal{M}=2$ and $14$ for $\rm SF_+$ and $\rm SF_-$ respectively.

\subsection{$\mathrm{SF}\boxtimes \mathrm{SF}$}
\label{sec:sf2}

As mentioned in Sec.~\ref{sec:add_prop}, Eqs.~\eqref{eq:Lambda-relation1} and \eqref{eq:Lambda-relation2} are only enough to determine $\hat n_v$ and $\hat{p}_v$. However, to determine $n_{v}$ and $p_v$, one usually needs additional constraints. In this section, we discuss an example that additional constraints are needed. 

This example is a double semion-fermion topological order $\rm SF \boxtimes SF$. It contains eight anyons, 
\begin{equation}
\mathcal{C}_{\rm SF\boxtimes SF} = \{1,f\}\boxtimes \{1,s_1\}\boxtimes\{1,s_2\},
\end{equation}
where $s_1$ and $s_2$ are two semions. Note that when stacking two $\rm SF$'s into $\rm SF \boxtimes SF$, the transparent fermions from each layer should be identified. Fusion rules and topological spins can be easily deduced from those of a single SF topological order. The permutation $\rho_m$ is given by
\begin{align}
1\leftrightarrow 1, \ f\leftrightarrow f,\ s_1 \leftrightarrow s_1 f, \ s_2 \leftrightarrow s_2 f, \nonumber\\
s_1s_2\leftrightarrow s_1s_2, \ s_1s_2f\leftrightarrow s_1s_2f,
\nonumber
\end{align}
The mirror eigenvalues are $\mu_{s_1}= \kappa_1$, $\mu_{s_2}=\kappa_2$,  $\mu_{s_1s_2}=\kappa_1\kappa_2$, and others can be obtained by composition.

It is insufficient to determine $\{n_v,p_v\}$ by Eqs.~\eqref{eq:Lambda-relation1} and \eqref{eq:Lambda-relation2} only when there are non-Majorana vortices in $\mathcal{B}$. So, we consider the Abelian $\Z_2^f$-gauged theories of $\mathcal{C}_{\rm SF \boxtimes SF}$:
\begin{equation}
\mathcal{B} =\{1, f, \tau, \tau f\} \boxtimes \{1, s_1\}\boxtimes \{1,s_2\}, \nonumber
\end{equation}
where $\theta_\tau = e^{i2\pi z /8}$. The chiral central charge $c= z+2 \modulo{8}$ and total quantum dimension $D_\mathcal{B}=4$. Taking $a=\{1, s_1, s_2, s_1s_2\}$ and $v=\{\tau, \tau s_1, \tau s_2, \tau s_1s_2\}$, we have the matrix $\Lambda_{a,v}$:
\begin{equation}
\Lambda = \frac{1}{2}\left(
\begin{matrix}
1 & 1 & 1 & 1 \\
1 & -1 & 1 & -1\\
1 & 1 & -1 & -1\\
1 & -1 & -1 & 1
\end{matrix}
\right).
\label{eq:SF2-Lambda}
\end{equation}
Substituting it into \eqref{eq:Lambda-relation1} together with $\{\mu_a\}$ and $\{\xi_a\}$, we obtain
\begin{align}
\frac{1}{2}\left(
\begin{matrix}
(1+\kappa_1)(1+\kappa_2)\\
(1-\kappa_1)(1+\kappa_2)\\
(1+\kappa_1)(1-\kappa_2)\\
(1-\kappa_1)(1-\kappa_2)
\end{matrix}
\right)
=
d_x\left(
\begin{matrix}
\hat n_\tau\\
\hat n_{\tau s_1}\\
\hat n_{\tau s_2}\\
\hat n_{\tau s_1s_2}
\end{matrix}
\right), \nonumber  \\
\frac{1}{2}\left(
\begin{matrix}
(1-\kappa_1)(1-\kappa_2)\\
(1+\kappa_1)(1-\kappa_2)\\
(1-\kappa_1)(1+\kappa_2)\\
(1+\kappa_1)(1+\kappa_2)
\end{matrix}
\right)
=
d_y\left(
\begin{matrix}
\hat p_\tau\\
\hat p_{\tau s_1}\\
\hat p_{\tau s_2}\\
\hat p_{\tau s_1s_2}
\end{matrix}
\right) . 
\label{eq:np_SF2}
\end{align}
Accordingly, $d_x=d_y=1$ in order to have integer solutions for $\hat{n}_v$ and $\hat{p}_v$. 

Let us consider the case that $\kappa_1=\kappa_2=+1$. Substituting this into \eqref{eq:np_SF2}, we have $\hat{n}_{\tau}=\hat{p}_{\tau s_1 s_2}=2$, and others are zero. That is,
\begin{align}
n_{\tau} + n_{\tau f} =2, \nonumber\\
p_{\tau s_1 s_2} + p_{\tau s_1 s_2 f} = 2.
\label{eq:SF2_n}
\end{align}
Then, $l(x^+)$ contains either $X_{\tau}^+$, or $X_{\tau f}^+$, or both. At this stage, it is already enough to determine the topological spin of $x^+$, which is given by
\begin{equation}
\theta_{x^+} = e^{i2\pi c/16}\sqrt{\theta_\tau} = e^{i2\pi(2c-2)/16}.
\end{equation}
Then, the anomaly indicator is given by $\eta_\mathcal{M} = i$, i.e., it lives on the surface of a $\nu=4$ bulk TCSC. The analysis is similar for other values of $\kappa_1$ and $\kappa_2$.

However, from Eq.~\eqref{eq:SF2_n}, we do not know the precise values of $n_\tau, n_{\tau f}$ and $p_{\tau s_1s_2}, p_{\tau s_1 s_2 f}$. The properties in \eqref{eq:hatnp-prop} do not help either. We need to look for other constraints. Here we consider a constraint from \eqref{eq:rr=r}, by taking $\alpha = X_\tau^+$ and $\beta = (s_1,s_1)^+$. They follow the fusion rule
\begin{equation}
X_\tau^+ \times (s_1,s_1)^+ = X_\tau^{q},
\end{equation}
where $q=\pm$ and the precise value is not important (though it can be determined if one follows Ref.~\onlinecite{folding}). Their restriction maps  are
\begin{align}
r\{X_\tau^+\} & = n_\tau x^+ + n_{\tau f} x_f^+  +\dots \nonumber\\
r\{(s_1, s_1)^+\} & = f^+ ,
\end{align}
where we have used properties in \eqref{eq:hatnp-prop}.  In order to satisfy \eqref{eq:rr=r}, we find that $n_\tau $ and $n_{\tau f}$ must be equal. Accordingly,
\begin{equation}
n_\tau = n_{\tau f} =1 .
\end{equation}
Similarly, one can show that $p_{\tau s_1 s_2} = p_{\tau s_1 s_2 f} =1 $ by considering $\alpha = X_{\tau s_1 s_2}$ and $\beta = (s_1, s_1)^+$ in \eqref{eq:rr=r}.

\subsection{$SO(3)_3$ topological order}
\begin{table}
	\caption{SET data, lifting coefficients as well as some other data for $\rm SO(3)_{3}^\kappa$ with $\kappa = \pm 1$.}
	\label{tab:so3}
	\begin{tabular}{c|cccc}
		\multicolumn{5}{c}{SET data}\\
		\hline 
		& $\ 1\ $ & $\ f\ $ & $\ s\ $ & $\ sf\ $ \\
		\hline 
		$\ \mu_a\ $ & $1$ & $-1$ & $\kappa$ & $-\kappa$ \\
		$\xi_a$ & $1$ & $1$ & $-1$ & $-1$\\
		\hline 
		
	\end{tabular}
	\hspace{15pt}
	
	\bigskip
	\vspace{0pt}
	\begin{tabular}{c|ccc}
		\multicolumn{4}{c}{$\kappa=+1$; $d_{x}=\sqrt{2}$, $d_{y}=1$}\\		
		\hline
		& $\ v_{1}\ $ & $\ v_{1}f\ $ & $\ v_{2} \ $  \\
		\hline 
		$\sigma_v$ & $1$ & $1$ & $\sqrt{2}$ \\
		$  n_v$ & $\ 1\ $ & $\ 1\ $ & $\ 0\ $   \\
		$  p_v$ & $\ 0\ $ & $\ 0\ $ & $\ 1 $  \\
		\hline 
	\end{tabular}\hspace{15pt}
	\begin{tabular}{c|ccc}
		\multicolumn{4}{c}{$\kappa=-1$; $d_{x}=1$, $d_{y}=\sqrt{2}$}\\	
		\hline
		& $\ v_{1}\ $ & $\ v_{1}f\ $ & $\ v_{2} \ $  \\
		\hline 
		$\sigma_v$ & $1$ & $1$ & $\sqrt{2}$ \\
		$  n_v$ & $\ 0\ $ & $\ 0\ $ & $\ 1\ $   \\
		$  p_v$ & $\ 1\ $ & $\ 1\ $ & $\ 0\ $  \\
		\hline 
	\end{tabular}
\end{table}

Next, we consider the simplest non-Abelian fermionic topological order, the $SO(3)_3$ topological order\cite{fidkowski13}.  Similarly to the SF topological order, it also contains four anyons
\begin{equation}
\mathcal{C}_{SO(3)_3} = \{1, f, s, sf\},
\end{equation}
where $s$ is again a semion with $\theta_s = i$. However, now $s$ is non-Abelian and obeys the fusion rule $s\times s= 1 + s + sf $. Other fusion rules and topological data can be found in Refs.~\onlinecite{fidkowski13}. Mirror permutation on the anyons is the same as in SF, given by Eq.~\eqref{eq:sf-permutation}. There are two possible mirror fractionalizations, with $\mu_{s}=\kappa=\pm1$. The two mirror-enriched topological orders are referred as $SO(3)_3^+$ and $SO(3)_3^-$, respectively. The values of $\{\mu_{a}\}$ and $\{\xi_{a}\}$ in $SO(3)_3^\pm$ are summarized in Table \ref{tab:so3}. 

Again, there are 16 possible $\mathbb{Z}_{2}^{f}$-gauged theories. They all lead to similar results. Let us focus on one of the possibilities, the $SU(2)_6$ topological order. It contains 7 anyons
\begin{equation}
\mathcal{B}=\{1,f,s,sf,v_{1},v_{1}f,v_{2}\},
\end{equation}
where $v_{1}$ and $v_{1}f$ are non-Majorana vortices and $v_{2}$ is a Majorana vortex. The topological spins are $\theta_{v_{1}}=\theta_{v_{1}f}=e^{i3\pi/16}$ and $\theta_{v_{2}}=e^{i15\pi/16}$, and the central charge is $c=9/4$. We refer the readers to Ref.~\onlinecite{fidkowski13} and references therein for other topological data, such as fusion rules and $S$ matrix.

Now, we determine the topological order $\mathcal{U}$.
The matrix $\Lambda_{a,v}$ is defined through Eq.~\eqref{eq:lambda_def} using the $S$ matrix of $\mathcal{B}$. Taking $a=\{1,s\}$ and $v=\{v_{1},v_{2}\}$, we obtain the same unitary $\Lambda$ matrix as the one in Eq.~\eqref{eq:SF-Lambda}. Applying \eqref{eq:Lambda-relation1} with $\mu_a$, $\xi_a$, and $\sigma_v$ in Table \ref{tab:so3}, we obtain 
\begin{align}
\left(\frac{1+\kappa}{\sqrt{2}}, 1-\kappa\right) & = (\hat{n}_{v_{1}},\hat{n}_{v_{2}}) d_x, \nonumber\\
\left(\frac{1-\kappa}{\sqrt{2}}, 1+\kappa\right) & = (\hat{p}_{v_{1}},\hat{p}_{v_{2}}) d_y.
\end{align}
To solve the equations with integer $\hat n$ and $\hat{p}$, we consider $\kappa=+1$ and $\kappa = -1$ separately:  
\begin{enumerate}
	\item if $\kappa=+1$, then $d_{x}=\sqrt{2}$, $d_{y}=1$, $\hat{n}_{v_{1}}=1$, $\hat{n}_{v_{2}}=0$, $\hat{p}_{v_{1}}=0$, $\hat{p}_{v_{2}}=2$.
	\item if $\kappa=-1$, then $d_{x}=1$, $d_{y}=\sqrt{2}$, $\hat{n}_{v_{1}}=0$,  $\hat{n}_{v_{2}}=2$, $\hat{p}_{v_{1}}=1$, $\hat{p}_{v_2}=0$.
\end{enumerate}
Compared with the quantum dimensions in Table \ref{tab:ito}, we find that $\mathcal{U}$ is $\mathrm{GiTO_{12}}$ if $\kappa=+1$, and $\mathrm{GiTO_{12}^{\prime}}$ if $\kappa = -1$.

For $\kappa=+1$,  we have $n_{v_1} = n_{v_1f}=p_{v_2}=1$ using Eqs.\eqref{eq:hatnp-prop}, \eqref{eq:hatn-def} and \eqref{eq:hatp-def}. Then, from Eq.~\eqref{eq:gl}, we obtain the following lifting maps 
\begin{equation}
\begin{split}
l(1^+) & = (1,1)^{+}+(f,f)^{-}+[s,sf],\\
l(f^+) & = (s,s)^{\kappa}+(sf,sf)^{-\kappa}+[1,f],\\
l(x^{+})&=X_{v_{1}}^{+}+X_{v_{1}f}^{+},\\
l(y^{+})&=X_{v_2}^{+},\\
\end{split}
\end{equation}
For $\kappa=-1$, the last two lifting maps are different. Again using Eqs.\eqref{eq:hatnp-prop}, \eqref{eq:hatn-def} and \eqref{eq:hatp-def},  we have $n_{v_2} = p_{v_1}=p_{v_1f}=1$ and the following lifting maps
\begin{equation}
\begin{split}
l(x^{+})&=X_{v_2}^{+},\\
l(y^{+})&=X_{v_{1}}^{+}+X_{v_{1}f}^{+}.\\
\end{split}
\end{equation}
The lifting coefficients are summarized in Table \ref{tab:so3}. The anomaly indicator $\eta_{\mathcal{M}}$ can be obtained by Eq.~\eqref{eq:etam-alt}. We obtain
\begin{equation}
\eta_{\mathcal{M}}=e^{i\kappa 3\pi/8}.
\end{equation}
It shows that $ SO(3)_3^{+}$ lives on the surface of 3D TCSCs with an index $\nu=3$, and $ SO(3)_3^{-}$ lives on the surface of 3D TCSCs with an index $\nu=13$. This agrees, and should agree, with Eq.~\eqref{eq:etam}.

There are many other mirror SETs that our method can be applied for. For example, the time-reversal symmetric (T-Pfaffian)$_\pm$ theories discussed in Refs.~\cite{chen14a,bonderson13, metlitski14} can be easily adapted into mirror SETs. We have checked with our method that the mirror version of (T-Pfaffian)$_+$ is anomaly-free, while (T-Pfaffian)$_-$ is anomalous with $\eta_\mathcal{M}=-1$, agreeing with the result of Ref.~\onlinecite{metlitski15a}. The calculation is straightforward and similar to the above, so we do not show it here.

\section{Conclusions}

\label{sec:conclusion}

In summary, we have explored general fermionic topological orders enriched by the mirror symmetry. We have followed and extended the folding approach, proposed in Ref.~\onlinecite{folding}, to fermion systems. In particular, we have derived the expression \eqref{eq:etam} for the mirror anomaly indicator $\eta_{\mathcal{M}}$. The derivation makes use of dimensional reduction and anyon condensation theory.  

Similarly to Ref.~\onlinecite{folding}, we have defined an alternative set of data $\{n_v,p_v\}$ through anyon condensation, in addition to the original ones $\{\mu_a\}$, to describe mirror symmetry fractionalization.  The two sets of data are ``dual'' to each other, as shown in relations \eqref{eq:Lambda-relation1} and \eqref{eq:Lambda-relation2}. However, compared to the results in the bosonic case\cite{folding}, our results are not as complete. We have not derived a (relatively) complete set of constraints on $\{n_v, p_v\}$. If we are able to do so, the constraints on $\{n_v,p_v\}$ can help to constrain possible values of $\{\mu_a\}$ and thereby provide  a better understanding of mirror SETs beyond mirror anomaly. One of the obstacles to derive these constraints is that we do not know the complete $\Z_2^f$-gauged theory $\mathcal{B}$ in general, but only know part of its properties such as the $\Lambda$ matrix. To study general properties of $\mathcal{B}$ is an interesting problem by itself. 

It is interesting to extend this approach to other anomaly indicators. For example, Ref.~\onlinecite{LapaPRB2019} obtained several anomaly indicators for the surface topological orders that live on the boundary of 3D bosonic and fermionic topological insulators, i.e., systems with $\mathcal{T}$ and a $U(1)$ symmetry group. Similar indicators should exist for 3D topological crystalline insulators, i.e., systems with $\mathcal{M}$ and $U(1)$. Also, one may study anomalies for systems with $\mathcal{M}$ and an internal symmetry group $G=\Z_2$, $SU(2)$, etc. We expect that our approach applies, after proper extensions, for deriving the mixed anomaly between $\mathcal{M}$ and $G$ (the anomaly solely due to $G$ cannot be obtained in this way). In principle, one can first gauge $G$ or a proper subgroup of $G$, then perform the folding trick and study the gapped domain by anyon condensation theory.  However, one may encounter difficulties in practice, e.g., the gauged SET is not known in general for an arbitrary $G$.  We leave these generalizations to future studies.

\begin{acknowledgments}
C.W. is grateful to Yang Qi and Chao-Ming Jian for a previous collaboration, on which this work heavily relies. We thank S. Q. Ning for very helpful discussions. This work was supported by Research Grant Council of Hong Kong (ECS 21301018).
\end{acknowledgments}

\appendix

\section{Some properties of $\Lambda$}
\label{sec:app_B_prop}

In this appendix, we prove two properties of the $\Lambda$ matrix, defined in Eq.~\eqref{eq:lambda_def}. This matrix is a block of the $S$ matrix of $\mathcal{B}$, up to some normalization. 

First, we would like to prove \eqref{eq:lambda_unitarity}, which we repeat here for convenience:
\begin{align}
\sum_{[a]}\Lambda_{a,v}\Lambda^*_{a,v'}  & = \delta_{[v],[v']} \label{eq:app_lambda0},\\
\sum_{[v]}\Lambda_{a,v}\Lambda^*_{a',v}  & = \delta_{a,a'} - \delta_{af,a'} 
\label{eq:app_lambda1},
\end{align}
where $a, a'\in\mathcal{C}$, and $v,v'\in\bar{\mathcal{C}}$, and the summation over $[a]$ ($[v]$) means that only one anyon in the pair $[a]$ ($[v]$) is summed over. To show the two relations, we notice that the $S$ matrix of $\mathcal{B}$ satisfies
\begin{align}
S_{af,b} & = S_{a,b}, \ S_{af,v} = -S_{a,v},\nonumber\\
S_{a,vf} & = S_{a,v}, \ S_{v',vf} = -S_{v', v},
\label{eq:app_lambda_s}
\end{align}
which can be easily obtained from \eqref{eq:monodromy1}, \eqref{eq:monodromy2} and the fact  that $f$ is Abelian. (If one of the two vortices $v$ and $v'$ is Majorana-type, the last equality shows that $S_{v,v'}=0$.) Since $S$ is a unitary and symmetric matrix, we have
\begin{align}
\delta_{v,v'} & = \sum_{\alpha\in\mathcal{B}} S_{\alpha,v} S_{\alpha,v'}^* \nonumber\\
&  = 2\sum_{[a]} S_{a,v} S_{a,v'}^* + \sum_{[w]} \frac{2}{\sigma_w^2} S_{w,v}S_{w,v'}^*, \label{eq:app_lambda2}\\
\delta_{vf,v'} & = \sum_{\alpha\in\mathcal{B}} S_{\alpha,vf} S_{\alpha,v'}^* \nonumber\\
&  = 2\sum_{[a]} S_{a,v} S_{a,v'}^* - \sum_{[w]} \frac{2}{\sigma_w^2} S_{w,v}S_{w,v'}^*. \label{eq:app_lambda3}
\end{align}
Adding up \eqref{eq:app_lambda2} and \eqref{eq:app_lambda3}, and making use of the definition \eqref{eq:lambda_def} of $\Lambda$, we have
\begin{equation}
\sum_{[a]} \Lambda_{a,v} \Lambda_{a,v'}^*  = \frac{1}{\sigma_v\sigma_{v'}}(\delta_{v,v'} + \delta_{vf,v'}) = \delta_{[v],[v']},
\end{equation}
which is exactly \eqref{eq:app_lambda0}. Similarly, we have
\begin{align}
\delta_{a,a'} & = \sum_{\beta\in\mathcal{B}} S_{a,\beta} S_{a',\beta}^* \nonumber\\
&  = 2\sum_{[b]} S_{a,b} S_{a',b}^* + \sum_{[v]} \frac{2}{\sigma_v^2} S_{a,v}S_{a',v}^*,\label{eq:app_lambda4} \\
\delta_{af,a'} & = \sum_{\beta\in\mathcal{B}} S_{af,\beta} S_{a',\beta}^* \nonumber\\
&  = 2\sum_{[b]} S_{a,b} S_{a',b}^* - \sum_{[v]} \frac{2}{\sigma_v^2} S_{a,v}S_{a',v}^*.\label{eq:app_lambda5}
\end{align}
Subtracting \eqref{eq:app_lambda4} with \eqref{eq:app_lambda5} and making use of the definition of $\Lambda$, we immediately obtain \eqref{eq:app_lambda1}.

Next, we show the following relation
\begin{equation}
 \sum_{[a]} \Lambda_{a,v}\Lambda_{\rho_m(a),v'} =\delta_{[v],\rho_m([v'])},
\label{eq:app_lambda6}
\end{equation}
where $a\in\mathcal{C}$ and $v,v'\in\bar{\mathcal{C}}$.  This relation holds only if $\mathcal{C}$ is mirror symmetric. To make sense of \eqref{eq:app_lambda6}, we need to define $\rho_m([v'])$, i.e., the action of mirror permutation on the pairs $\{[v]\}$.  Once it is defined, we will see that 
\begin{equation}
\Lambda_{\rho_m(a),v'} = \Lambda_{a, w}^*, \ \text{with } [w] = \rho_m([v']).
\label{eq:app_lambda9}
\end{equation}
Then, \eqref{eq:app_lambda6} follows immediately by combining \eqref{eq:app_lambda0} and \eqref{eq:app_lambda9}.  

To define the action of $\rho_m$ on the vortex pairs, we first consider the so-called Verlinde algebra $\mathsf{Ver}(\mathcal{C})$ associated with the fusion rules in $\mathcal{C}$. It is an algebra spanned by the elements $\mathbf{e}_a$ with $a\in\mathcal{C}$, which satisfy
\begin{equation}
\mathbf{e}_{a} \mathbf{e}_{b} = \sum_{c} N_{ab}^c \mathbf{e}_c, \quad \mathbf{e}^\dag_a = \mathbf{e}_{\bar{a}}.
\label{eq:ver}
\end{equation}
This is a commutative algebra since $N_{ab}^c=N_{ba}^c$. Therefore, all its irreducible representations are one dimensional, which are called fusion characters and denoted by $\lambda_j(a)$.  They satisfy
\begin{equation}
\lambda_j(a)\lambda_j(b) = \sum_c {N_{ab}^c}\lambda_j(c), \quad \lambda_j(\bar{a}) = \lambda_j^*(a).
\end{equation}
There are in total $|\mathcal{C}|$ fusion characters, labelled by $j=1, \dots, |\mathcal{C}|$, since the dimension of the algebra is $|\mathcal{C}|$. One can show that all fusion characters can be constructed from the $S$ matrix of $\mathcal{B}$ as follows
\begin{equation}
\lambda_j(a) = \frac{S_{a,\beta}}{S_{1,\beta}},
\label{eq:app_lambda7}
\end{equation}
where $j=j(\beta)$, i.e., each $\beta\in\mathcal{B}$ can be mapped to a fusion character $j(\beta)$.  The fact that \eqref{eq:app_lambda7} is a fusion character follows from the Verlinde formula \cite{kitaev2006} and the fact that Verlinde algebra $\mathsf{Ver}(\mathcal{C})$ is a subalgebra of the Verlinde algebra $\mathsf{Ver}(\mathcal{B})$.   It is not hard to see that $\beta$ and $\beta f$ are mapped to the same fusion character, due to properties in \eqref{eq:app_lambda_s}. One can show that there exists a one-to-one mapping between the pair $[\beta]$ and fusion character index $j$ \cite{Wang_unpub_2020}. 

In the presence of mirror symmetry, the fusion multiplicity satisfies $N_{ab}^c = N_{\rho_m(a)\rho_m(b)}^{\rho_m(c)}$. Then, we have  
\begin{align}
\lambda_j(\rho_m(a))\lambda_j(\rho_m(b)) & = \sum_c {N_{\rho_m(a)\rho_m(b)}^{\rho_m(c)}}\lambda_j(\rho_m(c)) \nonumber\\
& = \sum_c {N_{ab}^c}\lambda_j(\rho_m(c)), 
\end{align}
Accordingly, $\lambda_j(\rho_m(a))$ is also a fusion character of the Verlinde algebra $\mathsf{Ver}(\mathcal{C})$. Then, there must be an index $\tilde{j}$, such that $\lambda_{\tilde{j}}(a) = \lambda_j^*(\rho_m(a))$. In the case that $j=j(b)$, one can show that $\tilde{j} = j(\rho_m(b))$, consistent with the original action of $\rho_m$ on $\mathcal{C}$. In the case that $j=j(v)$, the action of $\rho_m$ on $v$ is not defined yet. We make use of the one-to-one mapping between $j$ and $[\beta]$ to define such an action: given $v$, there must exist $\tilde{v}$ such that
\begin{equation}
\frac{S_{a,\tilde{v}}}{S_{1,\tilde{v}}} =  \frac{S_{\rho_m(a),v}^*}{S_{1, v}} ,
\label{eq:app_lambda8}
\end{equation}
where we explicitly used \eqref{eq:app_lambda7}. As discussed above, $\tilde{v}$ is not unique, but $[\tilde{v}]$ is. So, we define 
\begin{equation}
\rho_m([v]) \equiv [\tilde{v}].
\end{equation}
Inserting the definition \eqref{eq:lambda_def} of $\Lambda$ into \eqref{eq:app_lambda8} and using the relation $S_{1,\beta} = d_\beta/D_{\mathcal{B}}$,  we obtain
\begin{equation}
\frac{\sigma_{\tilde{v}}}{d_{\tilde{v}}} \frac{d_v}{\sigma_v} \Lambda_{a, \tilde{v}} = \Lambda_{\rho_m(a), v}^*.
\end{equation}
Note that while we  can only determine $[\tilde{v}]$ through \eqref{eq:app_lambda8}, $\sigma_{\tilde{v}}$ and $d_{\tilde{v}}$ are the same for either anyon in $[\tilde{v}]$. Given $v$ and $\tilde v$, both $\{\Lambda_{a,\tilde v}\}_{a\in\mathcal{C}}$ and $\{\Lambda_{\rho_m(a), v}\}_{a\in\mathcal{C}}$ are vectors of length 2. Accordingly, we must have
\begin{equation}
\frac{d_v}{\sigma_v} = \frac{d_{\tilde{v}}}{\sigma_{\tilde{v}}}.
\label{eq:app_lambda10}
\end{equation}
Then, Eq.~\eqref{eq:app_lambda9} results, which further gives rise to \eqref{eq:app_lambda6}.

We comment that defining $\rho_m([v])$ extends the domain of $\rho_m$ from $\mathcal{C}$  to $\mathcal{B}$. However, this extension is different from the one discussed in Sec.~\ref{sec:gauging_z2f}. There, we define $\rho_m:\mathcal{B}_l\rightarrow \mathcal{B}_r$, which is an anti-equivalence between two topological orders. In this appendix, $\rho_m$ is extended to be a map between vortex pairs of a single topological order $\mathcal{B}$. In general, $\rho_m$ can not be further extended such that it becomes an anti-autoequivalence of $\mathcal{B}$.

\bibliography{lie.bib}

\end{document}